\documentclass{aa}  

\usepackage{graphicx}
\usepackage{xcolor}
\usepackage{tikz}
\usepackage{amsmath}
\usepackage{txfonts}
\usepackage{hyperref}
\hypersetup{colorlinks=true,linkcolor=blue,citecolor=blue,filecolor=blue,urlcolor=blue}
\usepackage{etoolbox}

\begin{document}

   \title{Binary companions of nearby supernova remnants found with Gaia}

   \titlerunning{Binary companions of nearby SNR}
   \authorrunning{D. Boubert et al.}

   \author{D. Boubert
          \inst{1},
          M. Fraser
          \inst{1,2},
          N. W. Evans
          \inst{1},
          D. A. Green
          \inst{3},
          \and
          R. G. Izzard \inst{1}
          }

   \institute{Institute of Astronomy, University of Cambridge, Madingley Rise, Cambridge, CB3 0HA, United Kingdom\\
              \email{\href{mailto:d.boubert@ast.cam.ac.uk}{d.boubert@ast.cam.ac.uk}, \href{mailto:nwe@ast.cam.ac.uk}{nwe@ast.cam.ac.uk}, \href{mailto:rgi@ast.cam.ac.uk}{rgi@ast.cam.ac.uk}}
             \and
             School of Physics, O’Brien Centre for Science North, University College Dublin, Belfield, Dublin 4, Ireland
             \\
             \email{\href{mailto:morgan.fraser@ucd.ie}{morgan.fraser@ucd.ie}} 
             \and
   Astrophysics Group, Cavendish Laboratory, 19 J. J. Thomson Avenue, Cambridge CB3 0HE, United Kingdom
\\
   	\email{\href{mailto:dag@mrao.cam.ac.uk}{dag@mrao.cam.ac.uk}}
   }

   \date{Received XXX; accepted YYY}

 
  \abstract
   {}
   {We search for runaway former companions of the progenitors of
    nearby Galactic core-collapse supernova remnants (SNRs) in the Tycho-{\it Gaia}
    astrometric solution (TGAS).}
   {We look for candidates among a sample of
     ten SNRs with distances $\lesssim 2\;\mathrm{kpc}$, taking
     astrometry and $G$ magnitude from TGAS and $B,V$ magnitudes from
     the AAVSO Photometric All-Sky Survey (APASS). A simple method of
     tracking back stars and finding the closest point to the SNR
     centre is shown to have several failings when ranking
     candidates. In particular, it neglects our expectation that
     massive stars preferentially have massive companions. We evolve a
     grid of binary stars to exploit these covariances in the
     distribution of runaway star properties in colour -- magnitude --
     ejection velocity space. We construct an analytic model which
     predicts the properties of a runaway star, in which the model
     parameters are the location in the grid of progenitor binaries
     and the properties of the SNR. Using nested sampling we calculate
     the Bayesian evidence for each candidate to be the runaway and
     simultaneously constrain the properties of that runaway and of
     the SNR itself.}
     {We identify four likely runaway companions of the Cygnus Loop (G074.0$-08.5$),
     HB 21 (G089.0$+04.7$), S147 (G180.0$+01.7$) and the Monoceros Loop (G205.5$+00.5$). HD 37424 has previously been suggested as the companion of S147, however the other three stars are new candidates. The favoured companion of HB 21 is
     the Be star BD+50 3188 whose emission-line features could be
     explained by pre-supernova mass transfer from the primary. There
     is a small probability that the $2\;M_{\odot}$ candidate
     runaway TYC 2688-1556-1 associated with the Cygnus Loop is a
     hypervelocity star. If the Monoceros Loop is related to the
     on-going star formation in the Mon OB2 association, the
     progenitor of the Monoceros Loop is required to be more massive
     than $40\;M_{\odot}$ which is in tension with the posterior
     for our candidate runaway star HD 261393.}
   {}


   \maketitle
%

\section{Introduction}
\label{s1}

Supernovae (SNe) mark the deaths of stars. They can be divided into
two broad categories: core-collapse SNe from the gravitationally
powered explosion of massive stars \citep[e.g.][]{smartt_progenitors_2009},
and Type Ia SNe from the thermonuclear destruction of white dwarfs
\citep[e.g.][]{hillebrandt_type_2000}. In both cases, there has been
considerable interest in recent years in understanding their
progenitor systems for reasons as diverse as testing stellar
evolutionary models, improving their use as cosmological probes, and
understanding their role in driving galactic evolution.

The most direct method for studying the progenitors of supernovae is
to detect them in pre-explosion imaging. This is limited however to
the handful of cases where SNe have exploded in a nearby galaxy with
deep, high-resolution images. Furthermore, it is most suited for
studying core-collapse supernova progenitors
\citep{smartt_progenitors_2009} which are luminous supergiants
(although see \citealp{li_exclusion_2011} and \citealp{mccully_luminous_2014} for applications to Type Ia
SNe). Alternative techniques to infer core-collapse SN progenitor
properties using nucleosynthetic yields from late-time spectroscopy of
SNe \citep[e.g.][]{jerkstrand_nebular_2014}, or hydrodynamic estimates
of ejecta mass \citep{bersten_iptf13bvn:_2014} are always model
dependent. For Type Ia SNe, the spectroscopic and photometric
signatures of interaction between SN ejecta and a companion may be
used to constrain the progenitor system, but are relatively weak
effects \citep{maeda_signatures_2014}.

Another means to study SN progenitors is to search for their former
binary companions that have survived the explosion. At least 70\% of
massive stars are seen to be in binary systems
\citep[e.g.][]{sana_binary_2012}. Furthermore, in a handful of cases, a
surviving binary companion has been detected in deep imaging of
extragalactic core-collapse SNe
\citep{maund_disappearance_2009,folatelli_blue_2014}. Type Ia SNe
require a binary companion to explode \citep{hillebrandt_type_2000},
and may leave behind a detectable non-degenerate companion
\citep[e.g.][and references
  therin]{han_companion_2008,pan_search_2014,noda_brightness_2016}. For
both Type Ia and core-collapse SNe, the stellar parameters of a
surviving binary companion can constrain the evolutionary
status of the SN progenitor at the point of explosion
\citep{maund_disappearance_2009,bersten_type_2012}. A SN progenitor
companion may also be polluted with metals from the explosion
(\citealp{israelian_evidence_1999} and more recently \citealp{liu_interaction_2015}).

Several searches have already been made for runaway stars in Galactic
Type Ia SN remnants, most notably in Tycho's SN where a possible
candidate (designated Tycho G) has been claimed to be the former
binary companion
\citep{ruiz-lapuente_binary_2004,gonzalez_hernandez_chemical_2009,bedin_improved_2014}. This
association has since been disputed
\citep{kerzendorf_subaru_2009,kerzendorf_high-resolution_2013,xue_newly_2015}. Searches
within other Galactic remnants such as that of SN 1006
\citep{gonzalez_hernandez_no_2012} and Kepler's SN
\citep{kerzendorf_reconnaissance_2014} have failed to yield a companion,
while a non-degenerate companion has been almost completely ruled out
for SNR 0509$-67.5$ in the Large Magellanic Cloud
\citep{schaefer_absence_2012}.

Searches for companions to core-collapse SNe have mostly focussed on
runaway OB stars near SN remnants
\citep{blaauw_origin_1961,guseinov_searching_2005}. HD 37424, a main
sequence B star, has been proposed to be associated with the SN
remnant S147 \citep{dincel_discovery_2015}. The pulsar PSR
J0826+2637 has been suggested to share a common origin with the
runaway supergiant G0 star HIP 13962 \citep{tetzlaff_origin_2014}, although there is no identified SNR. In
the Large Magellanic Cloud, the fastest rotating O-star (VFTS102) has
been suggested to be a spun-up SN companion associated with the young
pulsar PSR J0537$-6910$ \citep{dufton_vlt-flames_2011}.

Recently \citet{kochanek_cas_2017} used Pan-STARRS1 photometry \citep{chambers_pan-starrs1_2016}, the \citet{green_three-dimensional_2015} dust-map and the NOMAD \citep{zacharias_vizier_2005} and HSOY \citep{altmann_hot_2017} proper motion catalogues to search for runaway former companions of the progenitors of the three most recent, local core-collapse SNe: the Crab, Cas A and SN 1987A. Based on a null detection of any reasonable candidates \citet{kochanek_cas_2017} put limits on the initial mass ratio $q=M_2/M_1\lesssim0.1$ for the nominal progenitor binary of these SNRs. \citet{kochanek_cas_2017} note that this limit implies a 90\% confidence upper limit on the $q\gtrsim0.1$ binary fraction at death of $f_{\mathrm{b}}<44\%$ in tension with observations of massive stars.

The reason to search for runaway companions of core-collapse
supernovae is that their presence or absence can be used to constrain
aspects of binary star evolution. These include mass transfer rates,
common-envelope evolution and the period and binary fraction
distributions. Runaway stars are interesting in their own right for
their dynamical properties with the fastest runaway stars being
unbound from the Milky Way.

In this work, we present a systematic search for SN-ejected binary
companions within the {\it Gaia} Data Release 1
\citep[DR1;][]{gaia_collaboration_gaia_2016}. Unfortunately, the Kepler SNR is too distant at around $6\;\mathrm{kpc}$ to
search for a companion using DR1, however ten other remnants including
S147 lie within our distance cut-off of $2\;\mathrm{kpc}$. The data sources for both the
runaways and the SNR are discussed in Section~\ref{sec:data}. We then
outline two approaches to this problem -- the first using purely
kinematic methods in Section~\ref{sec:simple}, the second exploiting
colour, magnitude and reddening together with the peculiar velocity of the
progenitor binary with a Bayesian framework in Section~\ref{sec:bayesgrid}. For
four SNRs (the Cygnus Loop, HB 21, S147 and the Monoceros Loop) we identify
likely runaway companions which are discussed in detail in
Section~\ref{sec:results}.

\section{Sources of data}
\label{sec:data}

The list of candidate stellar companions for each SNR is taken from a cross-match of
TGAS and APASS (Sec. \ref{sec:tgas}). There is no analogously uniform
catalogue for SNRs and so we conduct a literature review for each SNR
to establish plausible estimates for the central position, distance
and diameter, which we discuss in Section \ref{sec:snr} and in
Appendix \ref{sec:boutique}.

\subsection{Summary of stellar data}
\label{sec:tgas}

On the 14th September 2016 the {\it Gaia} Data Release 1 (GDR1) was made
publicly available \citep{gaia_collaboration_gaia_2016-1,gaia_collaboration_gaia_2016}. The primary astrometric component of the release
was the realisation of the Tycho-{\it Gaia} astrometric solution
(TGAS), theoretically developed by \citet{michalik_tycho-gaia_2015},
which provides positions, parallaxes and proper motions for the stars
in common between the GDR1 and \emph{Tycho-2} catalogues. At
$1\;\mathrm{kpc}$ the errors in the parallax from TGAS typically
exceed $30\%$. To constrain the distance of the candidates we
nearest-neighbour cross-match with the AAVSO Photometric All-Sky
Survey (APASS) DR9 to obtain the $B{-}V$ colour. {\it Gaia} Data Release 2
is anticipated in early 2018 and will contain the additional blue
$G_{\mathrm{BP}}$ and red $G_{\mathrm{RP}}$ magnitudes. Substituting
for $B{-}V$ with the $G_{\mathrm{BP}}{-}G_{\mathrm{RP}}$ colour will allow
our method to be applied using only data from {\it Gaia} DR2.

\subsection{Summary of individual SNRs}
\label{sec:snr}

We select SNRs that are closer than $2\;\mathrm{kpc}$, having stars
in our TGAS-APASS cross-match within the central 25\% of the SNR by radius,
and lying within the footprint of Pan-STARRS so we can use the 3D
dustmap of \citet{green_three-dimensional_2015}. The SNRs in our sample are typically older than $10\;\mathrm{kyr}$ and so will have swept up more mass from the ISM than was ejected, which makes it difficult to type them from observations of their ejecta. We can say that these SNRs are likely the remnants of core-collapse SNe since around $80\%$ of Galactic SNe are expected to be core-collapse SNe \citep[e.g.][]{mannucci_supernova_2005,li_nearby_2011}. Moreover, several are identified with regions of recent star formation (i.e. G205.5$+00.5$ with Mon OB2) or molecular clouds in OB associations (i.e. G089.0$+04.7$ with molecular clouds in Cyg OB7). The properties of this sample of ten SNRs are given in Table \ref{tab:snr}, where $N_{\mathrm{TGAS}}$ is the number of candidates found in TGAS and $N_{\mathrm{TGAS+APASS}}$ is the number remaining after the cross-match with APASS.

\begin{table*}[]
	\centering
	\caption{Assumed properties of the sample of supernova remnants where the errors on the distance are $1\sigma$ and are described in Appendix \ref{sec:boutique}.}
	\label{tab:snr}
	\begin{tabular}{llllllll}
		\hline \hline SNR         & Known as         & RA       & Dec    & Diameter (arcmin) & Distance (kpc)              & $N_{\mathrm{TGAS}}$ & $N_{\mathrm{TGAS+APASS}}$ \\ \hline
		G065.3$+05.7$  & ---              & 19:33:00 & +31:10 & 310$\times$240  & $0.77\pm0.2$           & 294               & 7                       \\
		G069.0$+02.7$  & CTB 80           & 19:53:20 & +32:55 & 80       & $1.5\pm0.5$            & 14                & 11                      \\
		G074.0$-08.5$  & Cygnus Loop      & 20:51:00 & +30:40 & 230$\times$160  & $0.54_{-0.08}^{+0.10}$ & 115               & 76                      \\
		G089.0$+04.7$  & HB 21            & 20:45:00 & +50:35 & 120$\times$90   & $1.7\pm0.5$            & 25                & 3                       \\
		G093.7$-00.2$  & CTB 104A, DA 551 & 21:29:20 & +50:50 & 80       & $1.5\pm0.2$            & 10                & 10                      \\
		G114.3$+00.3$  & ---              & 23:37:00 & +61:55 & 90$\times$55    & $0.7\pm0.35$           & 19                & 17                      \\
		G119.5$+10.2$ & CTA 1            & 00:06:40 & +72:45 & 90       & $1.4\pm0.3$            & 8                 & 7                       \\
		G160.9$+02.6$ & HB 9             & 05:01:00 & +46:40 & 140$\times$120  & $0.8\pm0.4$            & 19                & 18                     \\ 
		G180.0$-01.7$  & S147             & 05:39:00 & +27:50 & 180      & $1.30_{-0.16}^{+0.22}$ & 36                & 31                      \\
		G205.5$+00.5$  & Monoceros Loop   & 06:39:00 & +06:30 & 220      & $1.2\pm0.4$            & 53                & 47                      \\ \hline \hline
	\end{tabular}
\end{table*}

Establishing a potential association between a star and a SNR requires
us to demonstrate a spatial coincidence at around the time of the
supernova explosion. The relevant properties of each SNR are then the location
of the centre $(\alpha,\delta)_{\mathrm{SNR}}$, distance
$d_{\mathrm{SNR}}$, age $t_{\mathrm{SNR}}$, angular diameter
$\theta_{\mathrm{SNR}}$ and either the proper motion
$(\mu_{\alpha\ast},\mu_{\delta})_{\mathrm{SNR}}$ or peculiar velocity
$(v_{\mathrm{R}},v_{\mathrm{z}},v_{\phi})_{\mathrm{SNR}}$.

We take the \citet[known as
\emph{SNRcat}]{ferrand_census_2012} and \citet{green_catalogue_2014} catalogues as the primary
sources of SNR properties. We use the positions and angular
diameters from the detailed version of the Green catalogue that is
available online\footnote{Green D. A., 2014, `A Catalogue of Galactic
  Supernova Remnants (2014 May version)', Cavendish Laboratory,
  Cambridge, United Kingdom (available at
  \url{http://www.mrao.cam.ac.uk/surveys/snrs/}).}.
The distance to a SNR is usually uncertain and so we describe the origin of each distance in Appendix \ref{sec:boutique}.


We do not use estimates of the ages of SNRs because distance estimates to SNRs are degenerate with
the age, so these two measurements are not
independent. We thus conservatively assume that the
supernova must be older than $1\;\mathrm{kyr}$ and younger than
$150\;\mathrm{kyr}$. A younger supernova at $1\;\mathrm{kpc}$ would very likely
be in the historical record \citep{stephenson_historical_2002,green_historical_2003} and the shell of an older SNR would no longer be
detectable.

Determining the location of the centre of a SNR is usually not straightforward. The
standard method to obtain the centre is to calculate the centroid of
the projected structure of the SNR shell on the sky, but this position
can be obfuscated by various effects such as the interaction between the ejecta and the local ISM,
overlap between SNRs, and background objects misclassified as
belonging to the SNR. G074.0$-08.5$ (Cygnus Loop) is notable for its peculiarity with a substantial blowout region to the south of the primary spherical shell \citep[e.g.][]{fang_numerically_2017}. A naive calculation of the centroid for this SNR would result in a centre which is around $10\;\mathrm{arcmin}$ away from the centroid of the shell. We have verified that our results for G074.0$-08.5$ are robust to this level of systematic error. Some of our SNR central positions have
associated statistical errors, but because these estimates do not in
general account for systematics we instead use a more conservative
constraint. We adopt a prior for the true position of the SNR centre
which is a two-dimensional Gaussian with a FWHM given by
\begin{equation}
\theta'=\mathrm{max}(5',0.05\theta_{\mathrm{SNR}}).
\end{equation}
These values were chosen to attempt to balance the statistical and systematic errors which are present.

We assume that the progenitor system was a typical binary in the Milky
Way thin disk and so is moving with the rotational velocity of the
disk together with an additional peculiar motion. We sample a peculiar velocity from the
velocity dispersions of the thin disk and propagate it into a
heliocentric proper motion. We take the Sun to be at $R_{\sun}=8.5\;\mathrm{kpc}$
and the Milky Way's disk rotation speed to be $v_{\mathrm{disk}} = 240\;
\mathrm{km}\;\mathrm{s}^{-1}$ with a solar peculiar velocity of
$(U_{\sun},V_{\sun},W_{\sun})=(11.1,12.24,7.25)\;\mathrm{km}\;\mathrm{s}^{-1}$
\mbox{\citep{schonrich_local_2010}}. We neglect uncertainties in
these values since they are subdominant.

\begin{figure}[t]
	\includegraphics[scale=0.45,trim = 0mm 0mm 0mm 2mm, clip]{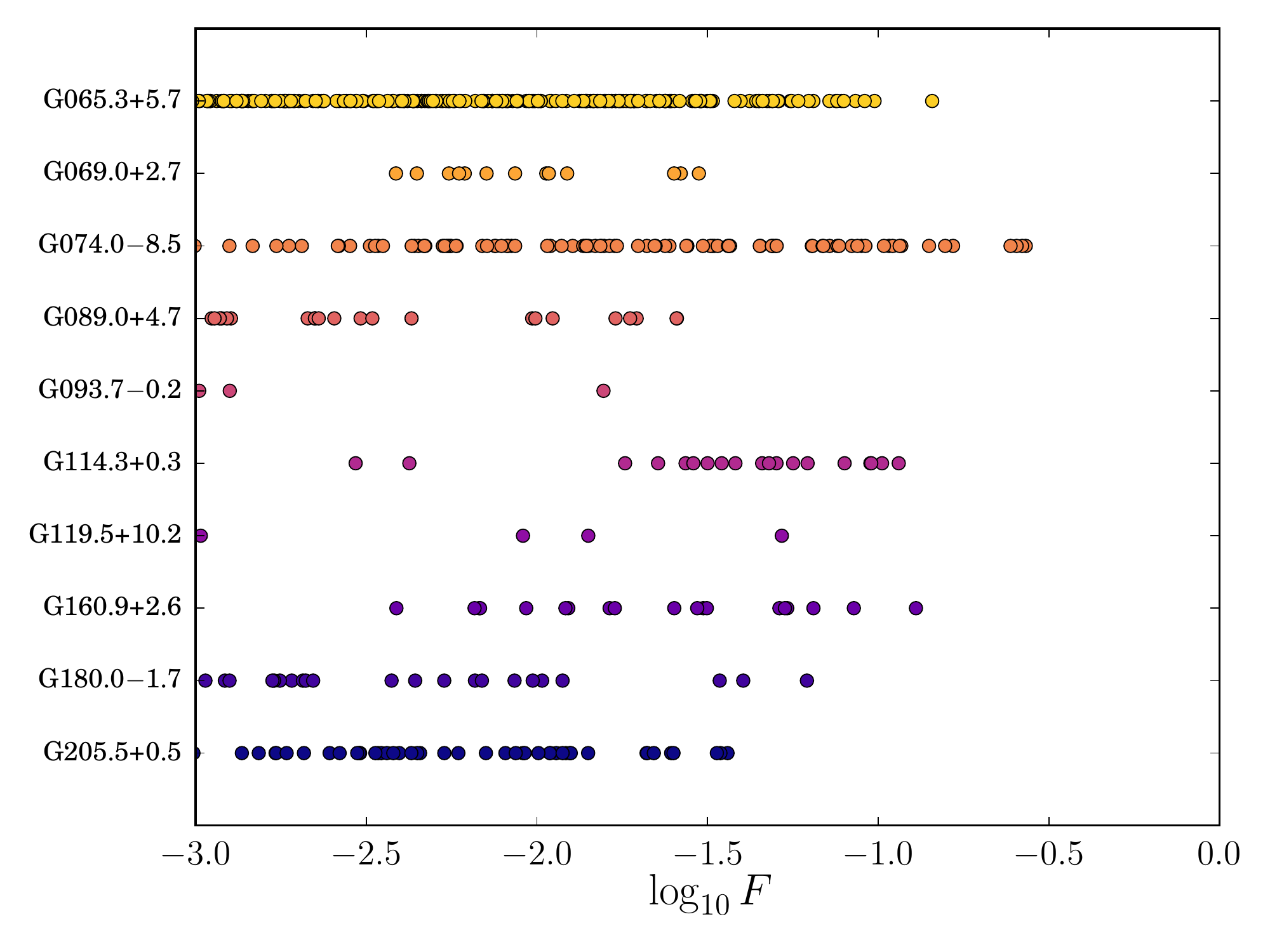}
	\caption{The fraction $F$ of realisations of each star in each SNR which are consistent with being spatially coincident with the centre of the SNR at one point in the past $150\;\mathrm{kyr}$. The criteria for determining whether a realisation is consistent are described in Section \ref{sec:simple}.}
	\label{fig:simple}
\end{figure}

\section{Searches with only kinematic constraints}
\label{sec:simple}

The typical expansion velocities of supernova remnant shells are more
than $1000\;\mathrm{km}\;\mathrm{s}^{-1}$ for the first few $10^4$
years of their evolution \citep{reynolds_supernova_2008}. Thus, since recent estimates for the maximum
velocity of runaways are $540\;\mathrm{km}\;\mathrm{s}^{-1}$ for late
B-types and $1050\;\mathrm{km}\;\mathrm{s}^{-1}$ for G/K-dwarfs
\citep{tauris_maximum_2015}, it is reasonable to assume that the former
companion to the SNR progenitor still resides in the SNR. For each
SNR we select all stars in TGAS that are within 25\% of the radius of the SNR
giving us somewhere in the range of $10\text{--}300$ stars per
SNR.  Less than 1\% of runaways are ejected with velocities in excess of $200\;\mathrm{km}\;\mathrm{s}^{-1}$ \citep[e.g.][]{eldridge_runaway_2011} thus considering every star in the SNR would increase the number of potential candidates by an order of magnitude while negligibly increasing the completeness of our search. Our choice to search the inner 25\% by radius is more conservative than the one sixth by radius searched by previous studies \citep[e.g.][]{guseinov_searching_2005,dincel_discovery_2015}. For each of these stars, we have positions $(\alpha,\delta)$,
parallax $\omega$ and proper motions $(\mu_{\alpha*},\mu_{\delta})$ as well
as the mean magnitude G, with a full covariance matrix $\mathrm{Cov}$ for the
astrometric parameters.

Given the geometric centre
$(\alpha_{\mathrm{SNR}},\delta_{\mathrm{SNR}})$ and proper motion
$(\mu_{\alpha,\mathrm{SNR}},\mu_{\mathrm{\delta,SNR}})$ of the
remnant and their errors, we can estimate the past location at
time $-t$ of each star by the equations of motion,
\begin{align}
\alpha_{*}(t)&=\alpha_{*}-t\mu_{\alpha\ast} \\
\delta(t)&=\delta-t\mu_{\delta},
\end{align}
and we can write similar expressions for the remnant centre. Note we use $*$ to denote quantities we have transformed to a flat space, for instance $\alpha_{*}=\alpha\cos\delta$. The angular separation $\Delta \theta$ is then approximated by,
\begin{equation}
\label{eq:theta}
\Delta\theta(t) = \sqrt{\left[\alpha_{*}(t)-\alpha_{*,\mathrm{SNR}}(t)\right]^2+\left[\delta(t)-\delta_{\mathrm{SNR}}(t)\right]^2}.
\end{equation}
Since the typical angular separations involved are less than a few degrees at all times this approximation is valid to first order. This expression has a clearly defined global minimum given by,
\begin{equation}
T_{\mathrm{min}}=\frac{(\alpha_{*}\!-\!\alpha_{*,\mathrm{SNR}})(\mu_{\alpha*}\!-\!\mu_{\alpha*,\mathrm{SNR}})+(\delta\!-\!\delta_{\mathrm{SNR}})(\mu_\delta\!-\!\mu_{\delta,\mathrm{SNR}})}{(\mu_{\alpha*}\!-\!\mu_{\alpha*,\mathrm{SNR}})^2+(\mu_\delta\!-\!\mu_{\delta,\mathrm{SNR}})^2},
\end{equation}
which can be substituted back into Equation \ref{eq:theta} to obtain the minimum separation $\Delta\theta_{\mathrm{min}}$.

We construct the covariance matrix
$\mathrm{Cov}=\mathrm{D}^{1/2}\mathrm{Corr}\mathrm{D}^{1/2}$ using the
correlation matrix $\mathrm{Corr}$ and the diagonal matrix of
errors
$\mathrm{D}=\mathrm{Diag}(\sigma_{\alpha}^2,\sigma_{\delta}^2,\sigma_{\omega}^2+(0.3\;\mathrm{mas})^2,\sigma_{\mu_{\alpha}}^2,\sigma_{\mu_{\delta}}^2)$ which are given in TGAS. We
have added on the $0.3\;\mathrm{mas}$ systematic error in parallax
recommended by \citet{gaia_collaboration_gaia_2016-1}. We draw samples
from the multivariate Gaussian distribution defined by the mean position
$(\alpha,\delta,\omega,\mu_{\alpha},\mu_{\delta})$ and the
covariance matrix $\mathrm{Cov}$ and from the distributions of the SNR
centre, distance and peculiar velocity. These latter distributions are
described in Section \ref{sec:snr}. We calculate $T_{\mathrm{min}}$
and $\theta_{\mathrm{min}}$ for each of the samples which can be combined into distributions for the predicted minimum separation and time
at which it occurs.

Once we have these distributions we classify stars by the
plausibility of them being the former companion. We do this
in a qualitative way by finding the fraction $F$ of realizations of each
star which satisfy $1<(T_{\mathrm{min}}/\mathrm{kyr})<150$, the
line-of-sight distance between the star and the SNR is less than
$153\;\mathrm{pc}$ and has $\theta_{\mathrm{min}}$ corresponding to a
physical separation less than $1\;\mathrm{pc}$. The latter two of
these constraints use the distance to the location of the progenitor
binary when the supernova exploded which can be calculated using the
sampled parameters and the time of the minimum separation. The $153\;\mathrm{pc}$
limit of the second constraint is simply the distance travelled by a star
at $1000\;\mathrm{km}\;\mathrm{s}^{-1}$ over $150\;\mathrm{kyr}$ and
is the maximum likely distance travelled by a runaway associated with
a SNR. The $1\;\mathrm{pc}$ limit of the third constraint is approximately the maximum likely separation of two stars in a binary and is smaller than the $153\;\mathrm{pc}$ limit in the radial direction because we have a measurement of the proper motion of each candidate. The value of $F$ is shown for every star in each SNR in Figure \ref{fig:simple}.

We rank the candidates in each SNR by this quasi-statistical measure and 
consider the star with the highest $F$ to be the most likely
candidate. For some of these stars, we can obtain APASS $B{-}V$
photometry from a TGAS/APASS cross-match and, were this method
effective, most of the best candidates would be blue. Of
the ten best candidates two have no associated $B{-}V$ in the
cross-match and five have $B{-}V>1.3$ hence are unlikely to be OB
stars. One of the two best candidates without a measured $B{-}V$ was HD 37424 in G180.0$-01.7$ (S147), which is one of the only five stars in G180.0$-01.7$ which did not have APASS magnitudes. HD 37424 has been previously suggested to be the runaway companion of G180.0$-01.7$ \citep{dincel_discovery_2015} and taking $B{-}V=0.073\pm0.025$ from that paper we see that our kinematic method would have proposed this star as a candidate if it had magnitudes in APASS. The remaining three stars with magnitudes are TYC 2688-1556-1 in G074.0$-08.5$ (Cygnus Loop) with
$B{-}V=0.43$, BD+50 3188 in G089.0$+04.7$ (HB 21) with $B{-}V=0.39$ and TYC
4280-562-1 in G114.3$+00.3$ with $B{-}V=0.39$. Of these stars only BD+50
3188 is specifically mentioned in the literature with
\citet{chojnowski_high-resolution_2015} concluding that it is a B star. That one of the stars is B type suggests that the other two stars with similar colour are also B type by association, although it is possible that these two stars are less reddened by interstellar dust. G089.0$+04.7$ is at a distance of $1.7\pm0.5\;\mathrm{kpc}$ while the other
SNRs are much closer at $0.54_{-0.08}^{+0.10}\;\mathrm{kpc}$ and
$0.70\pm0.35\;\mathrm{kpc}$. The consequence of the $B{-}V$ measurement being less reddened by dust is that the star is intrinsically redder and so the two untyped candidates may be A type or later.

This conclusion has some obvious problems. First, we have not
established what fraction of runaways from core-collapse SNe we would
expect to be later-type than OB. The distribution of mass ratios in massive binary systems is observed to be flat \citep[e.g.][]{sana_binary_2012,duchene_stellar_2013,kobulnicky_toward_2014} which suggests that we would expect most runaways from
core-collapse SNe to be bright, blue, OB-type stars. Thus, while it is
possible to have low-mass companions of primaries with masses
$M>8\;M_{\sun}$, around 80\% of companions to massive stars
will have masses in excess of $3\;M_{\sun}$ with a median
of $7\;M_{\sun}$. This expectation is in conflict with the result above where five out of the ten best candidates are likely to be low-mass stars. One explanation for this seemingly
large fraction of contaminants is that the method efficiently rules
out those stars which are travelling in entirely the wrong direction
to have originated in the centre of the SNR, but leaves in background
stars which are co-incident on the sky with the centre of the SNR and
whose proper motion is not constrained. This can explain the large
fraction of our best candidates being stars whose photometry indicates
that if they are at the distance of the SNR then they must be faint,
red, late-type stars, because more distant stars will be more dust-obscured and so appear redder.

A second problem is that, because we have not accounted for the
reddening in $E(B{-}V)$ in a quantitative way, the estimated spectral type of
our candidates depended on one of them having already been typed. This
estimated type is very uncertain, and two of the stars could be A type
or even later. The method we used also did not make use of the {\it
  Gaia} $G$ magnitude, despite it being the most accurate 
magnitude contained in the {\it Gaia}-APASS cross-match.  Third, while we have generated a list of candidates, the
ranking in the list is not on a firm statistical basis. There are four
stars in G074.0$-08.5$ which have $0.25<F<0.28$, only one of which is
our best candidate. It is difficult to defend a candidate when a
different statistical measure could prefer a different star.  The
fourth problem is that using the $B{-}V$ photometry to further constrain
our list of candidates relies on an expectation that most runaways from
core-collapse supernova should be OB stars. Ideally our statistical
measure should incorporate this prior but include the possibility that
some runaways will be late-type.

These problems point towards the need for an algorithm that
incorporates kinematics with photometry, dust maps and binary star
simulations in a Bayesian framework.

\section{Bayesian search with binaries, light and dust}
\label{sec:bayesgrid}

\subsection{Binary star evolution grid}
\label{sec:bingrid}

The three most important parameters which determine the evolution of a binary star are the initial primary mass $M_1$, initial secondary mass $M_2$ and initial orbital period $P_{\mathrm{orb}}$. Empirical probability distributions have been determined for these parameters and combining these with a model for binary evolution allows us to calculate a probability distribution for the properties of runaway stars. The properties of runaway stars which we are interested in are the ejection
velocity $v_{\mathrm{ej}}$, the intrinsic colour $(B{-}V)_0$ and the
intrinsic {\it Gaia} G magnitude $G_0$ at the time of the supernova.

There are two standard formalisms used when evolving a large number of binary stars to evaluate the probability distribution for an outcome. The first is Monte Carlo-based and involves sampling initial properties from the distributions and evolving each sampled binary. In this approach the initial properties of the evolved binaries are clustered in the high-probability regions of the initial parameter space and low-probability regions which may have interesting outcomes might not be sampled at all. The other method is grid-based and selects binaries to evolve on a regularly-spaced grid across the parameter space. This grid divides the parameter space into discrete elements (voxels) and the probability of a binary having initial properties which lie in that voxel can be found by integrating the probability distributions over the voxel. This probability is assigned to the outcome of the evolution of the binary that was picked in that voxel. The probability distribution for the runaway properties can be determined by either method. In the Monte Carlo approach the resulting runaways from the binary evolution are samples from the probability distribution for runaway properties while in the grid approach the distribution can be obtained by summing the probabilities which were attached to the runaway from each evolved binary.

Our choice of a grid over a Monte Carlo approach was motivated by our
need to probe unusual areas of the parameter space. A Monte Carlo
approach would require a large number of samples to fully explore
these areas, while a grid approach gives us the location and
associated probability of each voxel that produces a runaway star as
well as the properties of the corresponding runaway star. These
probabilities and other properties are thus functions of this initial
grid.

We model the properties of stars ejected from binary systems in which
one component goes supernova using the {\sc binary\_c}
population-nucleosynthesis framework
\citep{izzard_new_2004,izzard_population_2006,izzard_population_2009}.
This code is based on the binary-star evolution ({\sc bse}) algorithm
of \citet{hurley_evolution_2002} expanded to incorporate nucleosynthesis,
wind-Roche-lobe-overflow \citep{abate_wind_2013,abate_modelling_2015},
stellar rotation \citep{de_mink_rotation_2013}, accurate stellar
lifetimes of massive stars \citep{schneider_ages_2014}, dynamical
effects from asymmetric supernovae \citep{tauris_runaway_1998}, an
improved algorithm describing the rate of Roche-lobe overflow
\citep{claeys_theoretical_2014}, and core-collapse supernovae
\citep{zapartas_delay-time_2017}. In particular, we take our black
hole remnant masses from \citet{spera_mass_2015} and use a fit to the
simulations of \citet{liu_interaction_2015} to determine the impulse
imparted by the supernova ejecta on the companion, both of which were
options previously implemented in {\sc binary\_c}. We use version
2.0pre22, SVN 4585. Grids of stars are modelled using the {\sc
  binary\_grid2} module to explore the single-star parameter space as
a function of stellar mass $M$, and the binary-star parameter space in
primary mass $M_{1}$, secondary mass $M_{2}$ and orbital period $P_{\mathrm{orb}}$.

We pre-compute this binary grid of 8,000,000 binaries with
primary mass $M_1$, mass ratio $q=M_2/M_1$ and orbital period $P_{\mathrm{orb}}$ having the ranges,
\begin{align}
8.0 \leq M_1 / M_{\sun} &\leq 80.0,\nonumber\\
0.1 \; M_{\sun}/M_1  \leq q &\leq 1,\\
-1.0 \leq \log_{10} (P_{\mathrm{orb}}/\mathrm{days}) &\leq 10.0.\nonumber
\end{align}
We assume the primary mass has the
\mbox{\cite{kroupa_variation_2001}} IMF,
\begin{equation}
N(M_1)\propto
\begin{cases}
M_1^{-0.3}, & \mathrm{if}\ 0.01<M_1/M_{\sun}<0.08, \\
M_1^{-1.3}, & \mathrm{if}\ 0.08<M_1/M_{\sun}<0.5, \\
M_1^{-2.3}, & \mathrm{if}\ 0.5<M_1/M_{\sun}<80.0, \\
0, & \mathrm{otherwise.}
\end{cases}
\end{equation}
We assume a flat mass-ratio distribution for each system over the
range $0.1\;M_{\sun}/M_1<q<1$. We use a hybrid period
distribution \citep{izzard_temp_2017} which gives the
period distribution as a function of primary mass and bridges the
log-normal distribution for low-mass stars
\citep{duquennoy_multiplicity_1991} and a power law
\citep{sana_binary_2012} distribution for
OB-type stars. The grid was set at solar metallicity to model recent
runaway stars from nearby SNRs.

\begin{figure}[t]
\includegraphics[scale=0.53,trim = 2mm 7mm 30mm 0mm, clip]{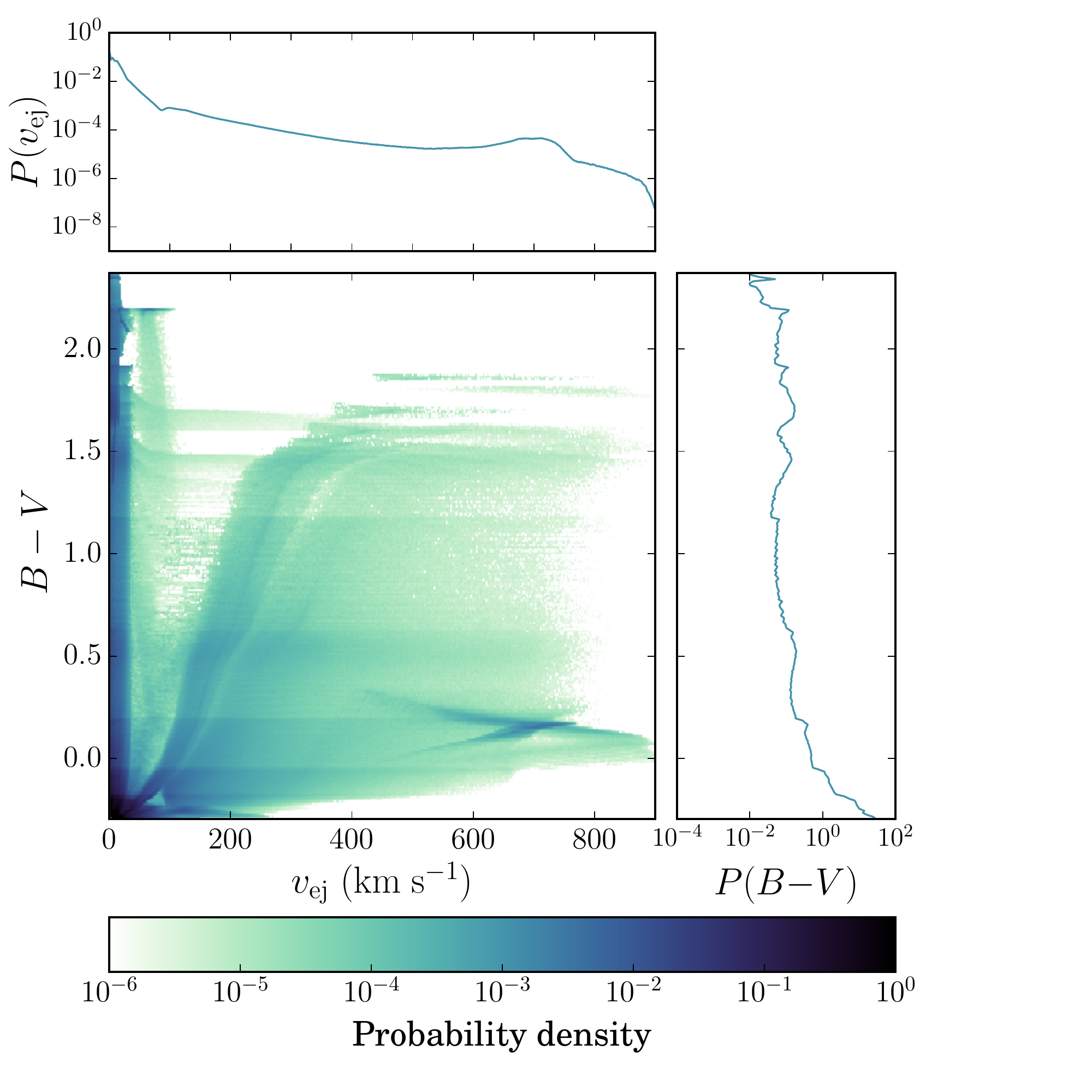}
\caption{Probability distribution in velocity-colour space of the
  runaways produced by our binary evolution grid. The top and right plots show 1D projections of the joint probability distribution.}
\label{fig:bvvej}
\end{figure}

It is useful to distinguish between the runaway parameter space
$(B{-}V)_0$--$G_0$--$v_{\mathrm{ej}}$, which is best for highlighting
the different runaway production channels, and the progenitor space
$M_1$--$q$--$P_{\mathrm{orb}}$, which is best for investigating
the connection of those channels to other binary phenomena. For
instance, our plot of runaway space in Figure \ref{fig:bvvej} has
several gaps towards the top right, which, when viewed instead in the
progenitor space, turn out to be regions where the binary has merged
prior to the primary going supernova. There are several prominent
trends in Figure \ref{fig:bvvej} which will be discussed in detail in
a forthcoming paper. We note that most of the probability is
concentrated on the left edge of the plot in slow runaways of
all colours. These correspond to the scenario of binary ejection where
the two stars do not interact and the ejection velocity is purely the
orbital velocity of the companion at the time of the supernova. The
rest of the structure corresponds to cases when at some point in the
evolution the primary overflows onto the secondary and forms a common
envelope \citep{izzard_common_2012,ivanova_common_2013}. The drag force of the gas on the two stellar cores causes an
in-spiral, while the lost orbital energy heats and ejects the common
envelope. These runaways are faster due to the larger orbital velocity
from the closer orbit, but there is a small additional kick from the
impact of SN ejecta on their surface.

\subsection{Algorithm}
\label{sec:algo}

We want to assess the hypothesis that a given observed star with
observables $\boldsymbol{x}$ is a runaway from a SNR. Our null hypothesis
$H_0$ is that a particular star is not the runaway companion
and we wish to test this against the hypothesis $H_1$ that it is. In
Bayesian inference each hypothesis $H$ has a set of model parameters
${\boldsymbol \theta}$ which can take values in the region $\mathit{\Omega}$. $H$ is
defined by a prior $\mathcal{P}({\boldsymbol \theta}|H)$ and a likelihood
$\mathcal{L}(\boldsymbol{x}|\mathbf{\boldsymbol \theta},H)$. The Bayesian evidence for
the hypothesis is then $\mathcal{Z}$, which is given by the integral
\begin{equation}
\mathcal{Z}=\int_{\mathit{\Omega}}\mathcal{P}({\boldsymbol \theta}|H)\; \mathcal{L}(\boldsymbol{x}|{\boldsymbol \theta},H) \;\mathrm{d}{\boldsymbol \theta}.
\label{eq:evidence}
\end{equation}
The evidence is equivalent to $\mathrm{Pr}(\boldsymbol{x}|H)$, i.e. the
probability of the data given the hypothesis.

To compare the background ($H_0$) and runaway ($H_1$) hypotheses we calculate the
Bayes factor $K=\mathcal{Z}_1/\mathcal{Z}_0$, where $\mathcal{Z}_0$
and $\mathcal{Z}_1$ are the evidences for $H_0$ and $H_1$
respectively. The interpretation of Bayes factors is subjective but a Bayes factor greater than one indicates that $H_1$ is
more strongly supported by the data than $H_0$ and vice versa. A review on the use of Bayes factors is given by \citet{kass_bayes_1995} who provide a table of approximate descriptions for the weight of evidence in favour of $H_1$ indicated by a Bayes factor $K$. To aid the interpretation of our results we replicate this table in Table \ref{tab:bayes}.

\begin{table}[]
	\centering
	\caption{Subjective interpretation of Bayes factors $K$ (taken from \citealp{kass_bayes_1995}).}
	\label{tab:bayes}
	\begin{tabular}{lll}
		\hline \hline $2\ln K$ & $K$       & Evidence against $H_0$             \\ \hline
		0 to 2   & 1 to 3    & Not worth more than a bare mention \\
		2 to 6   & 3 to 20   & Positive                           \\
		6 to 10  & 20 to 150 & Strong                             \\
		$>10$    & $>150$    & Very strong                     \\ \hline \hline
	\end{tabular}
\end{table}

To obtain the evidence for $H_0$ we define a probability distribution
using the stars in the TGAS/APASS cross-match that lie in an annulus
of width $10\degr$ outside the circle from which we draw our
candidates. Assuming that the locations of the stars in the space
$(\omega,\mu_{\alpha *},\mu_{\delta},G,B{-}V)$ can be described by a probability distribution we can approximate that distribution in a non-parametric way by placing Gaussians at the location of each star and summing up their contributions over the entire space. This method is called kernel density estimation (KDE). Note that we normalise the value in each
dimension by the standard deviation in that dimension for the entire
sample. This normalisation is necessary because the different
dimensions have different units. The prior for each candidate is a
Gaussian in each dimension centred on the measured value with a
standard deviation given by the measurement error. The likelihood for
a point sampled from the prior is the KDE evaluated at that
point. Strictly speaking this is the wrong way round. The KDE should
define the prior and the likelihood should be a series of Gaussians
centred on the data, but, since the definition of the evidence is
symmetric in the prior and likelihood (Eq. \ref{eq:evidence}), we are
free to switch them.


The evidence for $H_1$ is more complicated to calculate because the
model parameters ${\boldsymbol \theta}$ are properties of the SNR and
progenitor binary and thus need to be transformed into predicted
observables $\boldsymbol{\tilde{x}}$ of the runaway. The likelihood is
\begin{equation}
\mathcal{L}(\boldsymbol{x}|{\boldsymbol \theta})=\mathcal{N}(\boldsymbol{x}|\boldsymbol{\tilde{x}}({\boldsymbol \theta}),\mathrm{Cov}(\boldsymbol{x})),
\end{equation}
where $\boldsymbol{\tilde{x}}$ is a function of ${\boldsymbol \theta}$ and
$\mathcal{N}(\boldsymbol{a}|\boldsymbol{b},\boldsymbol{C})$ denotes the PDF of a
multivariate Gaussian distribution evaluated at $\boldsymbol{a}$ with mean
$\boldsymbol{b}$ and covariance matrix $\boldsymbol{C}$. For this preliminary
work we neglect the off-diagonal terms of the covariance matrix.

\begin{table}
\caption{Model parameters for the runaway hypothesis.}
\begin{tabular}{ l l }
\hline \hline
Parameter & Description \\ \hline
$\alpha_{\mathrm{SNR}}$ & RA of the true centre of the SNR \\
$\delta_{\mathrm{SNR}}$ & DEC of the true centre of the SNR \\
$d_{\mathrm{SNR}}$ & Distance to the true centre of the SNR \\
$t_{\mathrm{SNR}}$ & Age of the SNR \\
$M_1$ & Primary mass of the progenitor binary \\
$q=M_2/M_1$ & Mass ratio of the progenitor binary \\
$P_{\mathrm{orb}}$ & Orbital period of the progenitor binary \\
$E(B{-}V)$ & Reddening along the LoS to the candidate \\
$v_{\mathrm{R,pec}}$ & Peculiar velocity in Galactic $R$ \\ 
$v_{\mathrm{z,pec}}$ & Peculiar velocity in Galactic $z$ \\
$v_{\phi\mathrm{,pec}}$ & Peculiar velocity in Galactic $\phi$ \\\hline \hline
\end{tabular}
\label{tab:param}
\end{table}

The prior combines the primary mass $M_1$, mass ratio $q$ and
period $P_{\mathrm{orb}}$ of the progenitor binary with the location
($\alpha_{\mathrm{SNR}}$, $\delta_{\mathrm{SNR}}$), age
$t_{\mathrm{SNR}}$, distance $d_{\mathrm{SNR}}$ and peculiar velocity
$\boldsymbol{v}_{\mathrm{pec}}$ of the SNR and the reddening
$E(B{-}V)$ along the line of sight. These model parameters are
given in Table \ref{tab:param} for reference. The prior is
\begin{align}
\mathcal{P}({\boldsymbol \theta})=&\mathcal{N}(\alpha_{\mathrm{SNR}},\delta_{\mathrm{SNR}})\mathcal{N}(d_{\mathrm{SNR}})\mathcal{U}(t_{\mathrm{SNR}})P(M_1,q,P_{\mathrm{orb}}) \nonumber \\
&P(E(B{-}V)|d_{\mathrm{run}})\mathcal{N}(v_{\mathrm{R,pec}})\mathcal{N}(v_{\mathrm{z,pec}})\mathcal{N}(v_{\phi\mathrm{,pec}}),
\end{align}
where $\mathcal{N}(a)$ denotes a univariate Gaussian
distribution in $a$, $\mathcal{N}(a,b)$
denotes a multivariate Gaussian distribution in $a$ and
$b$, $\mathcal{U}(a)$ denotes a uniform distribution
in $a$ and the other components are non-analytic. The
additional variable $d_{\mathrm{run}}$ is the predicted distance
between the observer and runaway and is a function of the other model
parameters. The ranges, means and standard deviations for the first
three and last three distributions are given in Section \ref{sec:snr}
and were used for the simple method in Section \ref{sec:simple}.

The function $P(M_1,q,P_{\mathrm{orb}})$ is the
probability that, if there is a runaway star, it originates in a
progenitor binary with those properties. This probability can be
obtained directly from the PDFs of the binary properties (Section
\ref{sec:bingrid}) after renormalising to remove the binaries which do
not produce runaway stars. 

\begin{figure}[t]
	\includegraphics[scale=0.50,trim = 8mm 0mm 0mm 12mm, clip]{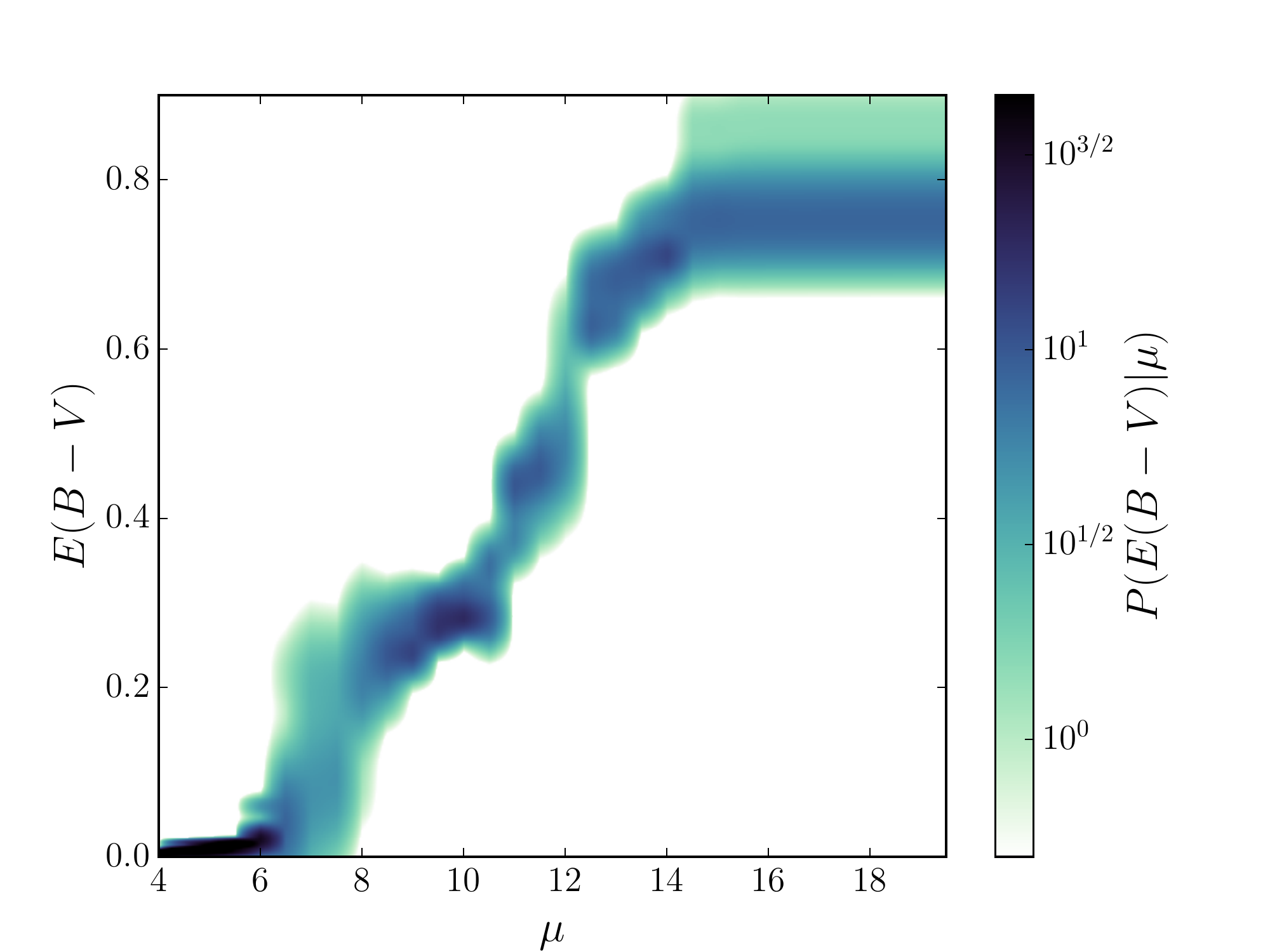}
	\caption{The conditional PDF for the dust extinction along the
          line of sight to G180.0$-1.7$ calculated by interpolating samples from the \citet{green_three-dimensional_2015} dust map.}
	\label{fig:ebv}
\end{figure}

The other non-analytic function
$P(E(B{-}V)|d_{\mathrm{run}})$ expresses the probability of the reddening along the line of sight to
the observed star if it is at a distance $d_{\mathrm{run}}$. \citet{green_three-dimensional_2015} used Pan-STARRS 1 and 2MASS
photometry to produce a 3D dustmap covering three quarters of the sky
and extending out to several
kiloparsec. \citet{green_three-dimensional_2015} provide samples from
their posterior for $E(B{-}V)$ in each distance modulus bin for each
HEALPix ($\mathtt{nside}=512$, corresponding to a resolution of approximately $7\;\mathrm{arcmin}$) on the sky. We use a Gaussian KDE to obtain a smooth probability distribution for
$E(B{-}V)$ in each distance modulus bin. We then interpolate between
those distributions to obtain a smooth estimate of $\mathrm{P}(E(B{-}V)|\mu)$
which we illustrate for one sight-line towards the centre of S147 in
Figure \ref{fig:ebv}. Note that $\log_{10}d_{\mathrm{run}}=1+\mu/5$,
where $d_{\mathrm{run}}$ is a function of our other model
parameters. \citet{green_three-dimensional_2015} used the same
definition of $E(B{-}V)$ as \citet{schlegel_maps_1998} so we have
converted their $E(B{-}V)$ to the Landolt filter system using
coefficients from \citet{schlafly_measuring_2011}.

The rest of this section is devoted to describing the transform
$\boldsymbol{\tilde{x}}({\boldsymbol \theta})$ between the model parameters and
the predicted observables. The outcome of the binary evolution is a
function solely of the progenitor binary model parameters. The
pre-calculated grid of binary stars thus provides the ejection
velocity $v_{\mathrm{ej}}$, intrinsic colour $(B{-}V)_0$ and intrinsic
magnitude $G_0$, which are essential to mapping the model parameters to
predicted observables. In addition, we obtain other parameters of
interest such as the present day mass of the runaway star
$M_{\mathrm{run}}$ and the age $T_{\mathrm{run}}$.

The kinematics of the SNR centre are fully determined by the position,
distance and peculiar velocity, under the assumption that the velocity
is composed of a peculiar velocity on top of the rotation of the
Galactic disk at the location of the SNR centre. The location of
the runaway on the sky is known because the errors on the observed
position of a star with {\it Gaia} are small enough to be
negligible. The remaining kinematics that need to be predicted are the
distance $d_{\mathrm{run}}$ and proper motion. The velocity vector of
the runaway $\boldsymbol{v}_{\mathrm{run}}$ is the sum of
the velocity of the SNR and the ejection velocity vector
$\boldsymbol{v}_{\mathrm{ej}}$. The location of the explosion, now the centre of the SNR, continues along the orbit of the progenitor binary within the Galaxy. We advance the centre of the SNR and the runaway along their
orbits for the current age of the SNR $t_{\mathrm{SNR}}$, noting that
this time is so short that any acceleration is negligible and thus the
orbits are essentially straight lines. The separation of the centre of the SNR and
the runaway at this point is then simply the difference of their
velocity vectors multiplied by $t_{\mathrm{SNR}}$,
i.e. $\boldsymbol{v}_{\mathrm{ej}}t_{\mathrm{SNR}}$. We then fix the
kinematics of the model by denoting the present-day centre of the SNR
to be at $(\alpha_{\mathrm{SNR}}, \delta_{\mathrm{SNR}})$ and the
present-day distance to the centre of the SNR to be
$d_{\mathrm{SNR}}$.

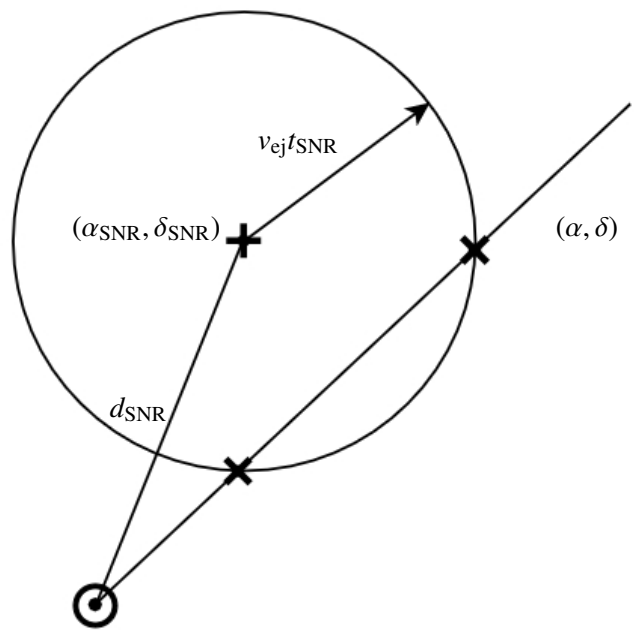
\begin{figure}[h]
	\begin{tikzpicture}
	\node[inner sep=0pt] (whitehead) at (0,0)
	{\includegraphics[scale=0.75,trim = 0mm 200mm 390mm 4mm, clip]{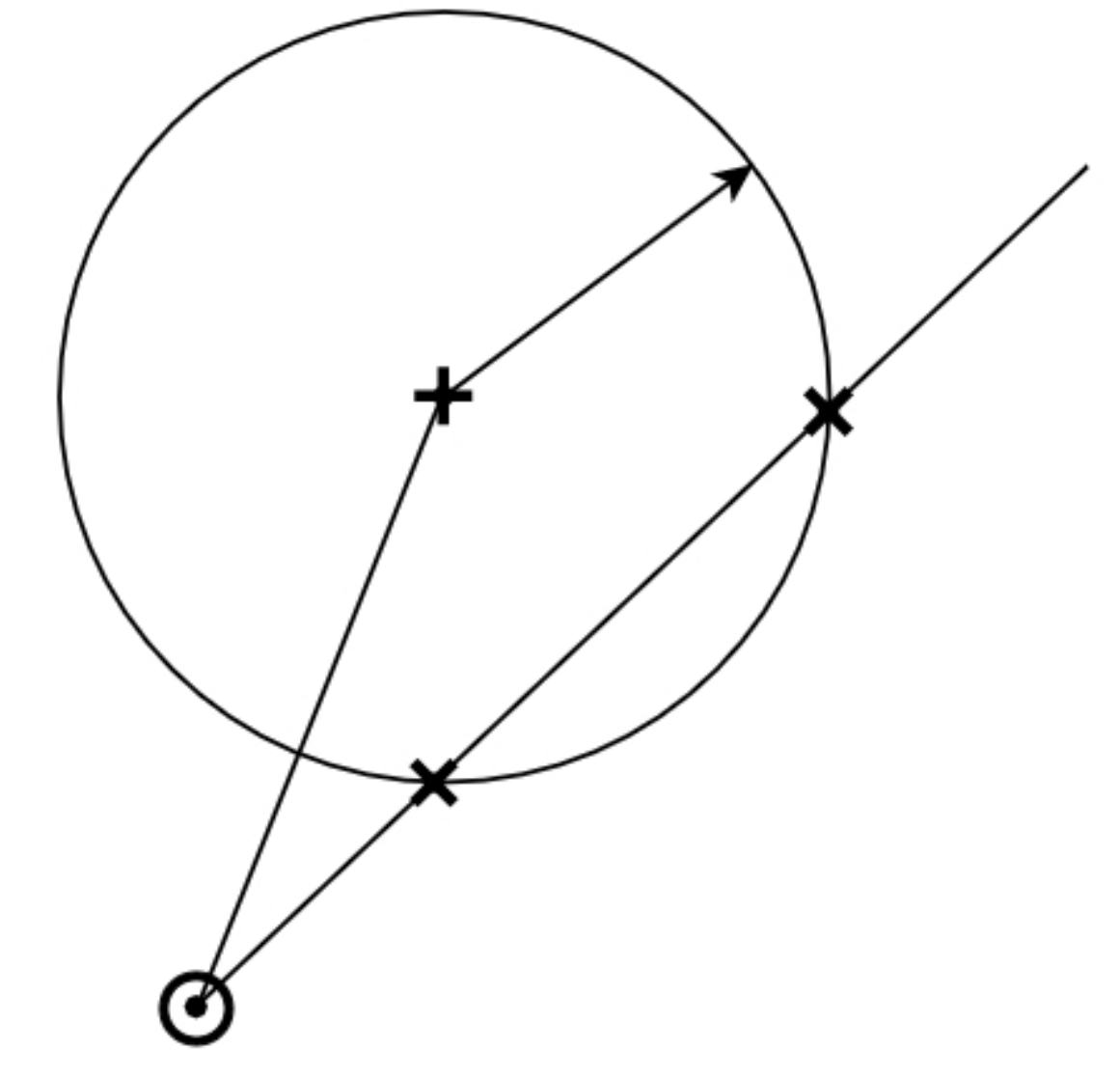}};
	\node[] at (-2.2,1.3) {\large$(\alpha_{\mathrm{SNR}},\delta_{\mathrm{SNR}})$};
	\node[] at (3.6,1.3) {\large$(\alpha,\delta)$};
	\node[] at (-0.2,2.4) {\large$v_{\mathrm{ej}}t_{\mathrm{SNR}}$};
	\node[] at (-2.3,-1.1) {\large$d_{\mathrm{SNR}}$};
	\end{tikzpicture}
	\caption{A diagram of the geometry described in Section
          \ref{sec:algo}. The observer is at {\large$\odot$} and the
          centre of the supernova remnant is at {\Large$+$}. The two
          possible locations of the candidate if it is the runaway are
          marked by {\large$\times$} and correspond to the
          intersection of the sphere of radius
          $v_{\mathrm{ej}}t_{\mathrm{SNR}}$ centred on the SNR and the
          half-line defined by the coordinates of the candidate.}
	\label{fig:geometry}
\end{figure}

To obtain predictions for the proper motions and parallax we consider
the intersection of the half-line defined by the observed position of
the candidate on the sky and a sphere centred at the distance and
position of the SNR. This sphere has a radius given by
$v_{\mathrm{ej}}t_{\mathrm{SNR}}$, which is the distance travelled by
the runaway since the supernova. A diagram of this geometry is shown
in Figure \ref{fig:geometry}. If the distance travelled by the runaway
is not large enough then the sphere fails to intersect the line and
thus the likelihood of this set of parameters is zero. In almost every case there are two intersections which correspond to the runaway
moving either away from or towards us. If the SNR is close and old and the
runaway is travelling rapidly, there is a pathological case in which
there is only one solution because the solution which corresponds to a
runaway moving towards us is already behind us. The geometry of the
intersection point gives us the distance to the star which we can use
to predict the parallax. The predicted proper motion of the runaway
depends on the velocity of progenitor binary. We sample in the
velocity dispersion of the Milky Way thin disk and add on the rotation
of the disk and ejection velocity of the runaway. This velocity is
converted to proper-motions and line-of-sight radial velocities using
the transforms of \citet{johnson_calculating_1987}.

We then obtain a prediction for $B{-}V$ by simply using $B{-}V=(B{-}V)_0 +
E(B{-}V)$. The {\it Gaia} $G$ band is broader than the $V$ band and so is more sensitive to the slope of the spectrum (Sanders et al., in preparation). One consequence of this is that the relative reddening in the $G$ band $A(G)/A(V)$ is a function of the intrinsic colour of the star.
Assuming that $A(V)=R_{V}E(B{-}V)$, where the constant $R_V=3.1$ is related to the average size of the dust grains and has been empirically determined in the Milky Way \citep{schultz_interstellar_1975}, we recast this dependency as
$A(G)/E(B{-}V)$ as a function of $(B{-}V)_0$. This relation has been calculated empirically by Sanders et al. (in preparation) and thus we have an expression for the apparent
magnitude $G=G_0+A(G)+\mu$.

We elected to use nested sampling \citep{skilling_nested_2006} to
explore the parameter space since it is optimised with estimating the evidence as the primary goal while more standard
Markov Chain Monte Carlo (MCMC) methods are targeted at obtaining samples from the posterior which afterwards can be used to estimate the evidence. We use the {\sc MultiNest}
implementation of nested sampling
\citep{feroz_multimodal_2008,feroz_multinest:_2009,feroz_importance_2013}
which we access through the {\sc PyMultiNest} {\sc Python} module
\citep{buchner_x-ray_2014}. {\sc MultiNest} requires that we express
our prior as a transform from a unit hypercube to the space covered by
our prior. For independent parameters, this is a trivial application
of inverting the cumulative distribution function. However, we have
two prior probability distributions $P(E(B{-}V)|\mu)$ and
$P(M_1,q,P_{\mathrm{orb}})$ for which there are no suitable
transforms. Note that $\mu$ is the distance modulus to the runaway
which is a complicated function of the position,
distance and age of the SNR and the ejection velocity of the
runaway. For these parameters, we use the standard method of moving
the probability distribution into the likelihood, which is implemented
in {\sc MultiNest} by assuming a uniform distribution in the prior and
including a factor in the likelihood to remove this extra
normalisation. Some technical details of the implementation of {\sc
  MultiNest} are discussed in Appendix~\ref{sec:multinest}.

Using nested sampling, we explore the parameter space and obtain a
value for the log of the evidence for each candidate. We then obtain
the Bayes factor by dividing the evidence for $H_1$ by the evidence
for $H_0$. A Bayes factor less than one indicates that the null hypothesis
is more strongly favoured, i.e. this star is likely a background star. A
Bayes factor greater than one suggests that the runaway model is
preferred.

\subsection{Fraction of supernovae with runaways}
\label{sec:runfrac}

Only a fraction of supernovae will result in a runaway companion. Some
massive stars are born single and companions are not always
gravitationally unbound from the compact remnant after the
supernova. Companions of massive stars tend to also be massive and so
some will themselves explode as a core-collapse supernova, either in a
bound system with the compact remnant of the primary or after being
ejected as a runaway star \citep[e.g.][]{zapartas_delay-time_2017}. A further contaminant is that binary
evolution can cause stars to merge before the primary supernova
occurs, through dynamical mass transfer leading to a spiral-in during
common envelope evolution. Our model assumes that there was a runaway
companion to the SNR and thus the calculated evidence needs to be
multiplied by the fraction of SNRs with a runaway.

Evolving a population of binary stars as described above we find that
the average number of core-collapse supernovae per binary system with
a primary more massive than $8\;M_{\sun}$ is 1.22. All single
stars in the mass range $8<M/M_{\odot}<40$ are expected to go supernova, with most stars more massive than $40\;M_{\odot}$ probably collapsing directly to black holes \citep{heger_how_2003}. Note that in the version of {\sc binary\_c} used for this work a
core-collapse supernova is signalled whenever the core of a star
collapses to a neutron star or black hole, including the case where the primary collapses directly to a black hole. Such collapses are
sufficiently rare that we do not correct for this effect. An
assumption on the binary fraction is required to combine statistics
for single and binary populations. \cite{arenou_simulated_2010}
provides an analytic empirical fit to the observed binary fraction of
various stellar masses,
\begin{equation}
F_{\mathrm{bin}}(M_1)=0.8388\tanh(0.079+0.688M_1).
\end{equation}
Based on this binary fraction and grids of single and binary stars evolved with {\sc binary\_c} we estimate that 32.5\% of core-collapse supernovae have a runaway
companion. This fraction is best described as `about a third' given
the approximate nature of the prescriptions used to model the binary evolution and the uncertainties in the empirical distributions of binary properties.

\subsection{Verification}
\label{sec:verification}

We verify our calculation of the evidence above by sampling runaways
from the model and using their 
$(\omega,\mu_{\alpha,\ast},\mu_{\delta},G, B{-}V)$ to generate a kernel
density estimate of their PDF. The evidence for a candidate to be a
runaway can then be computed identically to the background
evidence. In contrast to the method described in Section
\ref{sec:algo}, where the prior and likelihood are functions of the model parameters which are described in Table \ref{tab:param}, this
method casts the prior and likelihood as a function of the model observables. In the limit where we draw infinite samples from our model this method will give the same result as the method in Section \ref{sec:algo}. Drawing samples from the model and constructing a KDE is advantageous for its simplicity. The likelihood function is
an evaluation of a KDE and thus is guaranteed to be smooth and
non-zero everywhere, meaning that the considerations discussed in
Appendix~\ref{sec:multinest} are not relevant. The first disadvantage of
calculating the evidence by this method is that it only gives accurate
values of the evidence for regions of the parameter space which are
well sampled. The second disadvantage is that by not being explicit
about the model parameters we cannot directly constrain them,
and so this method does not output the maximum-likelihood
distance to the SNR or the mass of the progenitor primary. The implicit method is used in this work solely as a
cross-check of our results.

\subsection{Validation}
\label{sec:validation}

We validate our method by considering approximations to the false
positive and the false negative rate. For each SNR we assume there is
a nominal SNR at Galactic coordinates
$(l,b)=(l_{\mathrm{SNR}}-1\degr,b_{\mathrm{SNR}})$ with the same
distance and diameter estimates as the true SNR. We acquire candidates
from our TGAS and APASS cross-match and inject an equal number of
model runaway stars sampled from our binary grid, calculating
equatorial coordinates, parallaxes and proper motions which would
correspond to a runaway from that location ejected in a random
direction. These artificial measurements are convolved with a typical
covariance matrix of errors, here using the mean covariance matrix in
our list of candidates for this nominal SNR. For the dust correction,
we randomly select one of the twenty samples provided by the
\citet{green_three-dimensional_2015} dustmap in each distance modulus
bin along each line of sight, corresponding to the sight-line and
distance modulus that we have sampled for the runaway. The injected
runaways and real candidates are shuffled together so that the
algorithm described in Section \ref{sec:algo} is applied in the same manner to both the real stars and fake runaways. Since there is not a real SNR at this location all the real stars selected from the cross-match should be preferred to be background stars, while by construction the fake injected runaways should prefer the runaway hypothesis. An injected runaway which returns a Bayes factor $K<1$
is a false negative and a real star with $K>1$ is a false positive.

\begin{figure*}[t]
	\includegraphics[scale=0.48,trim = 4mm 6mm 3mm 0mm, clip]{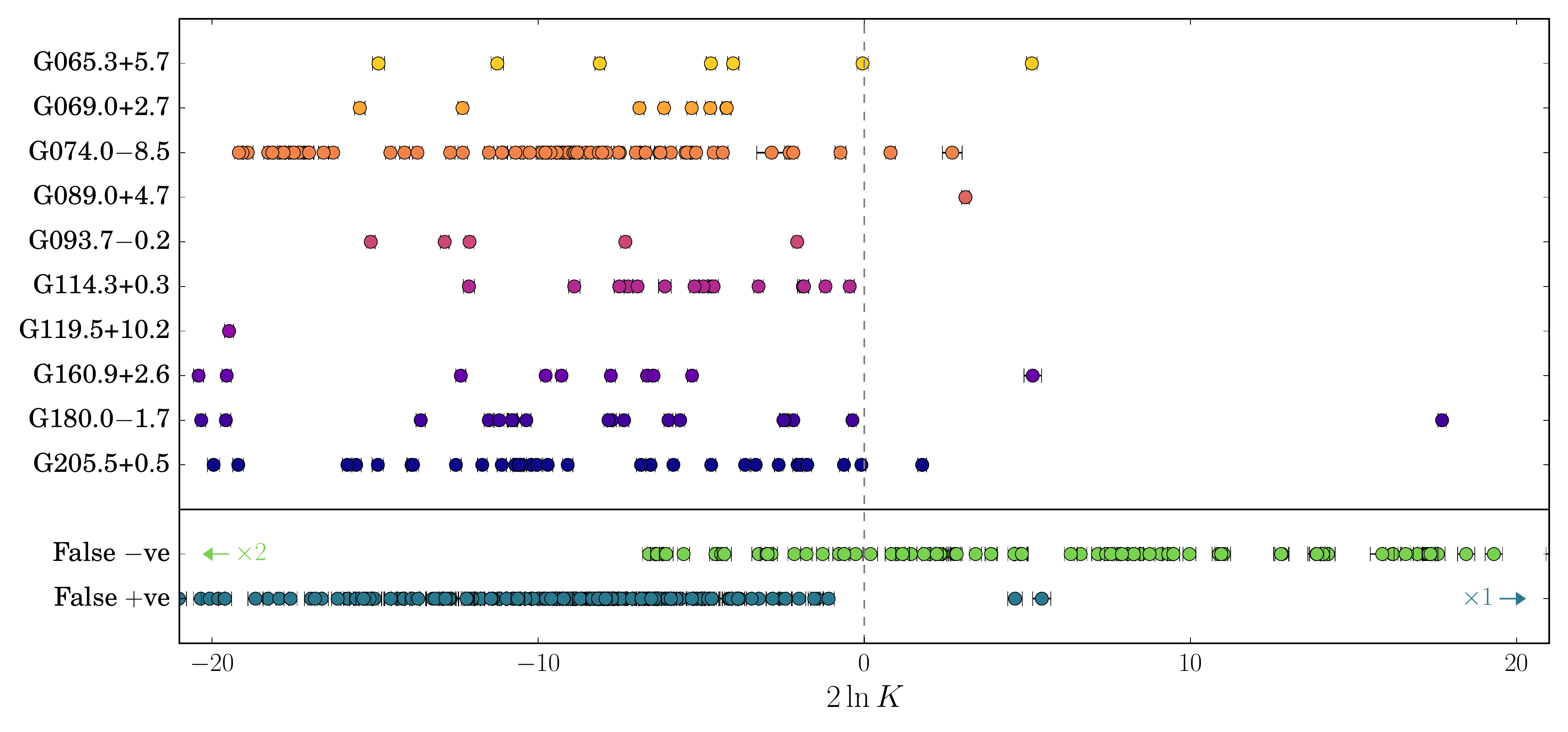}
	\caption{Bayes factors for the hypothesis that each star in each SNR is a
          runaway star versus the hypothesis that it is a
          contaminant. The false positive and false negative series are described in Section \ref{sec:validation}.}
	\label{fig:bayes}
\end{figure*}

At the bottom of Figure \ref{fig:bayes}, we show the calculated Bayes
factor $K$ for all the real stars and injected stars. There are 217
stars in each series. Only three of the real stars are returned as
false positives giving a false positive rate of 1.3$\%$. All three of
these false positives are from the fake version of G065.3$+05.7$ which
we find is because of the large photometric errors of APASS in
this field. These errors are around $\pm0.142$ in $B{-}V$ which compare
to $\pm0.055\;\mathrm{mag}$ for G180.0$-01.7$. This suggests that the
millimag precision of the $G_{\mathrm{BP}}$ and $G_{\mathrm{RP}}$
bands in {\it Gaia} DR2 will further reduce the false positive rate.

There are 22 false negatives which corresponds to a false negative
rate of about $10\%$. Given that we only expect a third of SNRs to
have an associated runaway companion (see Sec. \ref{sec:runfrac}) and
that we only consider ten SNRs, we should have at most one false negative in our observed sample.

We note that there are more stars closer to the $2\ln{K}=0$ boundary in our science runs (above the line in Fig. \ref{fig:bayes})
than were found in the false positive test. This is because runaways
are more likely to be OB stars and that OB stars are typically found
in star-forming regions. If a SN has occurred then a star-forming
region is nearby and so there are OB stars in or close to the SNR
which act as contaminants.

\section{Results}
\label{sec:results}

We report the seven stars for which the Bayes factor is
greater than one by at least the error on the evidence estimated by {\sc MultiNest}. In Figure
\ref{fig:bayes}, we show the calculated Bayes factor $K$ for the
candidates in each SNR over the range $(-20,20)$. In Section
\ref{sec:considerations} we discuss the three contaminant stars which
we are able to rule out and in Section \ref{sec:indivcand} we analyze
each of the four real candidates individually. Three of our candidates are new while HD 37424 in S147 has previously been suggested by \citet{dincel_discovery_2015}.

The only SNR in common with the search for OB runaways by
\citet{guseinov_searching_2005} is G089.0$+04.7$ and they proposed a
different candidate, GSC 03582$-00029$. This star appears to be
significantly brighter in the infrared ($J=10.2, H=9.7, K=9.6$)
than in the optical ($B=11.9$) while an OB star should have $B{-}K=-1$ \citep{castelli_new_2004}, so the classification of this star as
OB seems unlikely.

\begin{table}[]
	\centering
	\caption{Table of median posterior values for our new candidates.}
	\label{tab:candidates}
	\resizebox{0.48\textwidth}{!}{%
	\begin{tabular}{llll}
		\hline \hline
		SNR                                                 & G074.0$-08.5$      & G089.0$+04.7$     & G205.5$+00.5$     \\
		Candidate                                           & TYC 2688-1556-1   & BD+50 3188       & HD 261393        \\ \hline
		Sp. Type                                            & ---               & OB-              & B5V              \\
		$2\ln K$                                            & $0.81\pm0.15$     & $3.10\pm0.12$    & $1.78\pm0.14$    \\
		$\Delta\theta\;(\mathrm{arcmin})$                   & 11.45             & 2.43             & 8.54             \\
		$d_{\mathrm{SNR}}\;(\mathrm{kpc})$                  & $0.57\pm0.07$     & $1.69\pm0.26$    & $1.32\pm0.24$    \\
		$t_{\mathrm{SNR}}\;(\mathrm{kyr})$                  & $100.28\pm30.02$  & $107.17\pm27.77$ & $115.64\pm23.28$ \\
		$\log_{10}M_1\;(M_{\odot})$                & $1.11\pm0.17$     & $1.23\pm0.18$    & $1.07\pm0.12$    \\
		$q$                                                 & $0.16\pm0.05$     & $0.52\pm0.18$    & $0.44\pm0.11$    \\
		$\log_{10}P_{\mathrm{orb}}\;(\mathrm{days})$                       & $3.72\pm0.65$     & $2.41\pm1.36$    & $2.86\pm1.34$    \\
		$E(B{-}V)$                                            & $0.06\pm0.01$     & $0.55\pm0.08$    & $0.16\pm0.04$    \\
		$v_{\mathrm{pec}}\;(\mathrm{km}\;\mathrm{s}^{-1})$ & $29.20\pm10.34$   & $13.70\pm10.66$  & $16.14\pm20.41$  \\
		$(B{-}V)_0$                                           & $0.20\pm0.06$     & $-0.25\pm0.02$   & $-0.19\pm0.03$   \\
		$G_0$                                               & $2.39\pm0.27$     & $-2.52\pm0.47$   & $-0.92\pm0.49$   \\
		$v_{\mathrm{ej}}\;(\mathrm{km}\;\mathrm{s}^{-1})$   & $161.60\pm193.32$ & $32.25\pm17.73$  & $38.82\pm26.09$  \\
		$v_{\mathrm{r}}\;(\mathrm{km}\;\mathrm{s}^{-1})$    & $45.74\pm246.32$  & $-16.99\pm41.04$ & $23.82\pm51.94$  \\
		$M_{\mathrm{run}}\;(M_{\odot})$            & $1.73\pm0.13$     & $10.85\pm2.93$   & $5.78\pm1.15$ \\ \hline \hline  
	\end{tabular}}
\end{table}

\subsection{Eliminating the contaminants}
\label{sec:considerations}

In Figure \ref{fig:bayes}, there are six stars which have Bayes
factors greater than one. The presence of two of these stars for
G074.0$-08.5$ makes it clear that there is at least some level of
contamination. We found empirically that there are two ways to produce
false positives in our model. First, if the star is a high
proper-motion star in the foreground then the evidence for it in the
background model can be spuriously low. This can occur because the
background is constructed by taking a kernel density estimate of stars
around the SNR and it may not contain enough foreground stars to
reproduce this population. A low evidence in favour of the background
model boosts the Bayes factor so that the runaway model is preferred,
even if the star would be a very low-likelihood runaway. Second, if
the errors on the photometry from APASS are greater than around
$0.1\;\mathrm{mag}$ in each of $B$ and $V$ then it is possible for the
algorithm to ascribe a high probability to a far-away blue star when
the candidate is actually a nearby red star. This increases the
likelihood in favour of the runaway hypothesis.

If a contaminant is caused by the first of these possibilities, then
this is clear from an unusually jagged posterior of the runaway model. Foreground high proper-motion stars tend to not be OB
stars and so to explain the star under the runaway hypothesis {\sc
  MultiNest} is forced to sample in regions of the progenitor binary
parameter space that produce fast, red runaways. These are rare and lie in the region to the top right of Figure \ref{fig:bvvej} that is not well sampled in the binary grid because there are very few of them. This under-sampling results in
a jagged posterior dominated by spikes of high probability, with
reported modes that are poorly converged with large errors on
$\ln\mathcal{Z}$. The stars with the highest Bayes factor in both
G074.0$-08.5$ and G160.9$+02.6$ are contaminants of this first kind,
which can clearly be seen in Figure \ref{fig:bayes} as both these
stars have much broader error bars than the typical candidate.

The second type of contaminant is only a problem in this work because
we have chosen to take the photometry from APASS for all the SNRs,
while for some fields Tycho2 has much smaller errors. This is mainly
caused by a known problem in measurements taken for APASS DR8 in
Northern fields where the blue magnitudes have larger
errors than expected\footnote{\url{https://www.aavso.org/apass}}. If
the best measurement of $B{-}V$ has a large error then the problem
discussed above is a feature, because the Bayesian evidence is the
likelihood integrated against the probability of every possible
combination of model parameters. The star with the highest Bayes
factor in G065.3$+05.7$ is one such contaminant.  BD+30 3621 was the
only star in G065.3$+05.7$ with a Bayes factor greater than one. APASS
reports a measurement $(B{-}V)=1.10\pm0.88$ for this star, but {\sc
  MultiNest} picked out a most likely value of
$(B{-}V)_0=-0.22\pm0.02$. The Tycho 2 catalogue reports
$B{-}V=1.37\pm0.02$ confirming BD+30 3621 as a late-type star. The large
measurement error reported in APASS allowed the model to explore
parameter space where this star is much bluer than in reality. This is a preliminary study in preparation for {\it Gaia} DR2,
which will provide $G_{\mathrm{BP}}$ and $G_{\mathrm{RP}}$ with
millimag precision across the entire sky. This second type of
contaminant will not be a problem in {\it Gaia} DR2, because there
will not be more accurate photometry that we could use to
`double check' the measurements.

\begin{figure}[h]
	\includegraphics{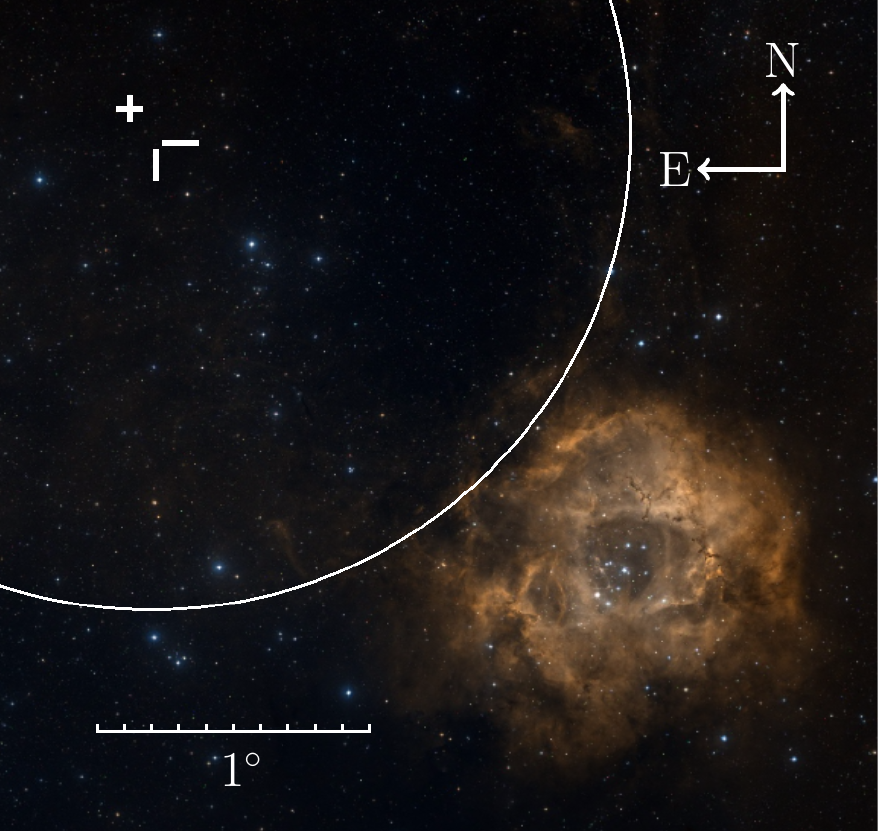}
	
	\caption{Digitized Sky Survey image of the vicinity of the
          Rosette Nebula. The runaway star candidate HD 261393 is marked by a white
          cross, the white cross hairs indicate the geometric centre
          of the Monoceros Loop and the white circle approximately
          shows the inner edge of the Monoceros Loop shell. The
          Rosette Nebula is in the bottom right and the Mon OB2
          association extends $3\degr$ to the east and north-east
          towards the centre of the Monoceros Loop.}
	\label{fig:rosette}
\end{figure}

\begin{figure*}[h]
	\includegraphics[scale=0.28,trim = 55mm 45mm 40mm 40mm, clip]{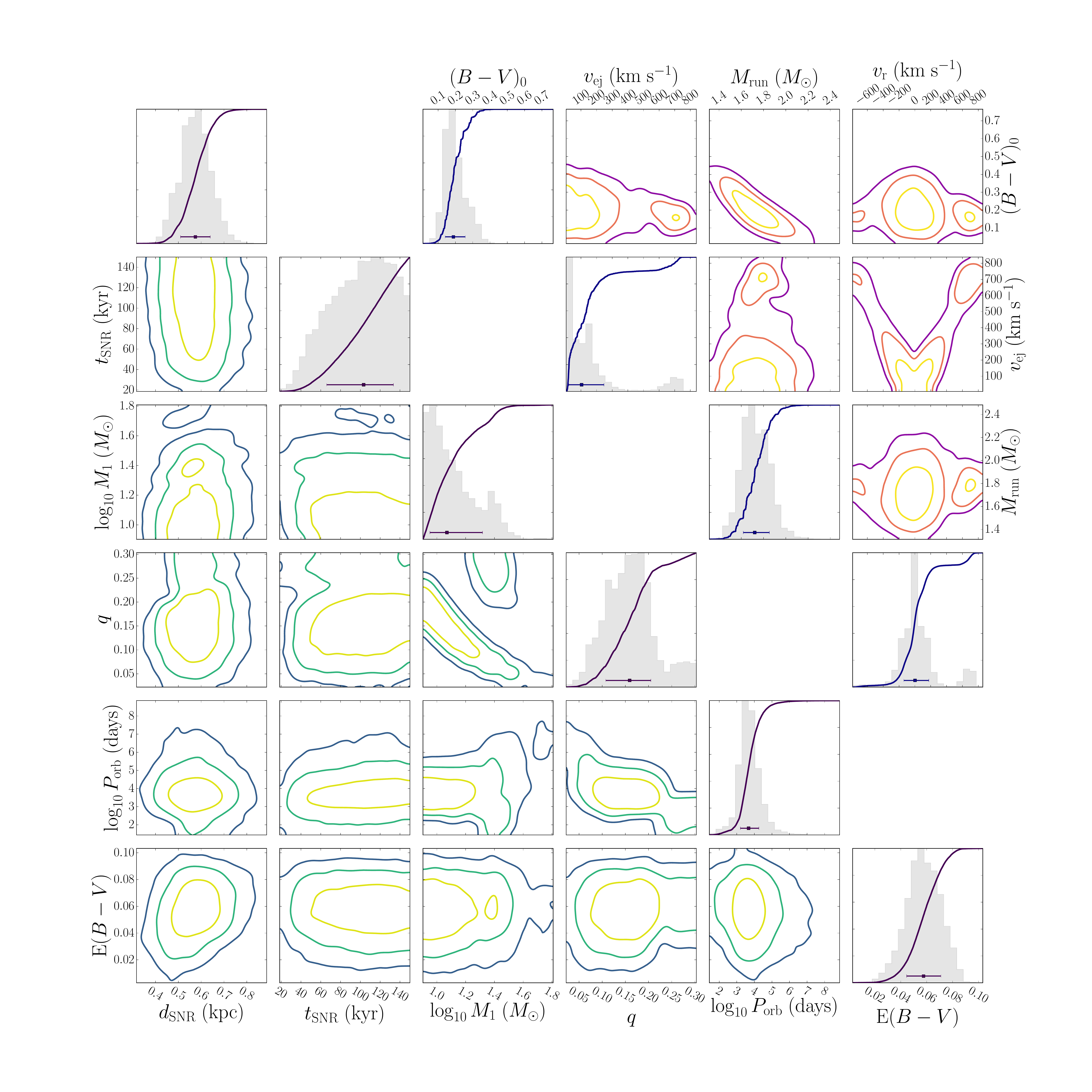}
	\caption{Corner plots of the posterior samples from the model
          of TYC 2688-1556-1 in G074.0$-08.5$ with $1\sigma$, $2\sigma$
          and $3\sigma$ contours. The 1D histograms include the CDF of
          that parameter and the error bars indicate the median and
          $1\sigma$ errorbars of each mode. \textbf{Bottom left:} A
          corner plot showing the model parameters, excluding the five
          parameters related to the position and peculiar velocity of
          the SNR which did not have covariances with the other
          parameters. \textbf{Top right:} A corner plot showing a
          selection of the derived parameters which are functions of
          the model parameters.}
	\label{fig:marginalG074.0}
\end{figure*}

\begin{figure*}[h]
	\includegraphics[scale=0.28,trim = 55mm 45mm 40mm 40mm, clip]{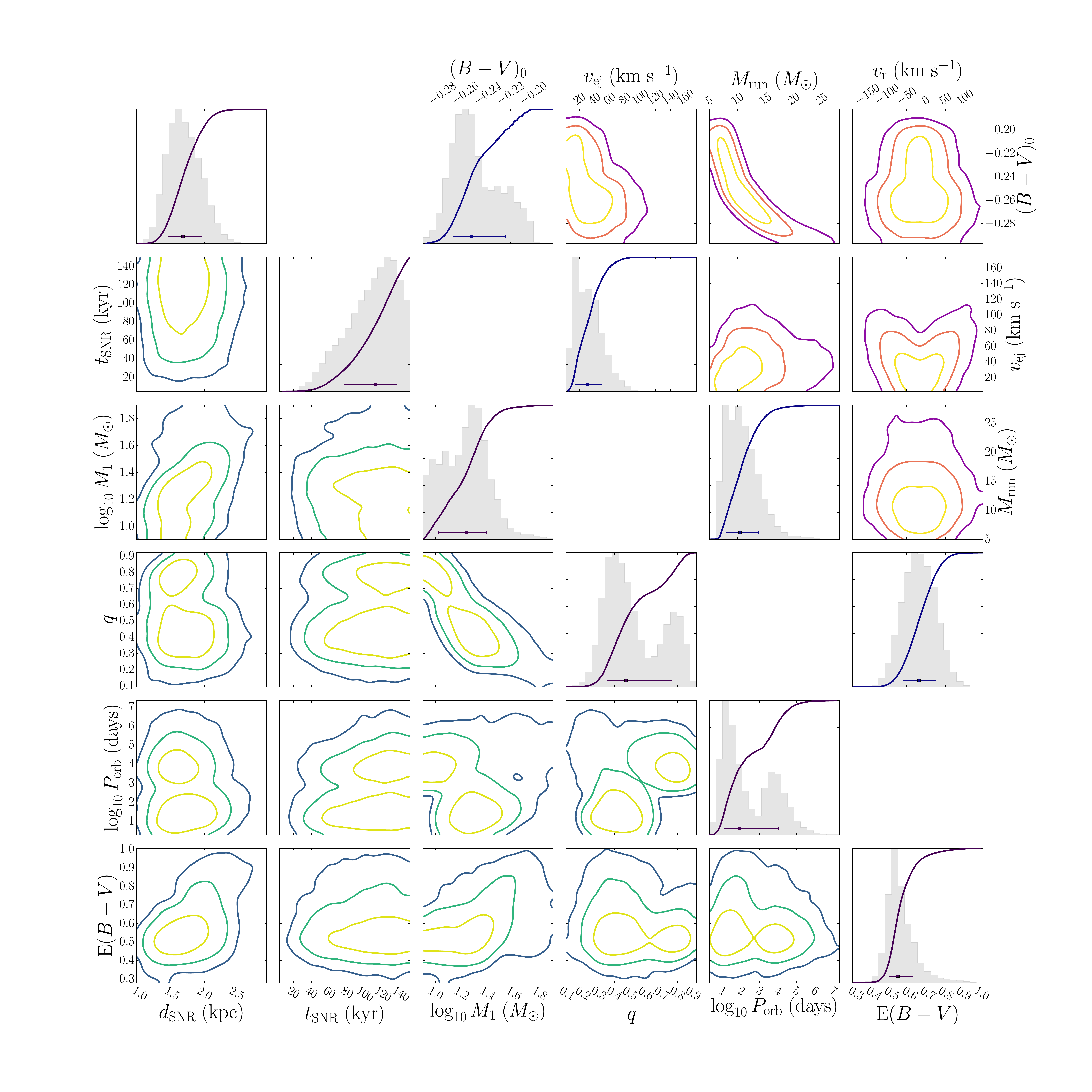}
	\caption{As Fig. \ref{fig:marginalG074.0} but for the
          candidate BD+50 3188 in G089.0$+04.7$.}
	\label{fig:marginalG089.0}
\end{figure*}

\begin{figure*}[h]
	\includegraphics[scale=0.28,trim = 55mm 45mm 40mm 40mm, clip]{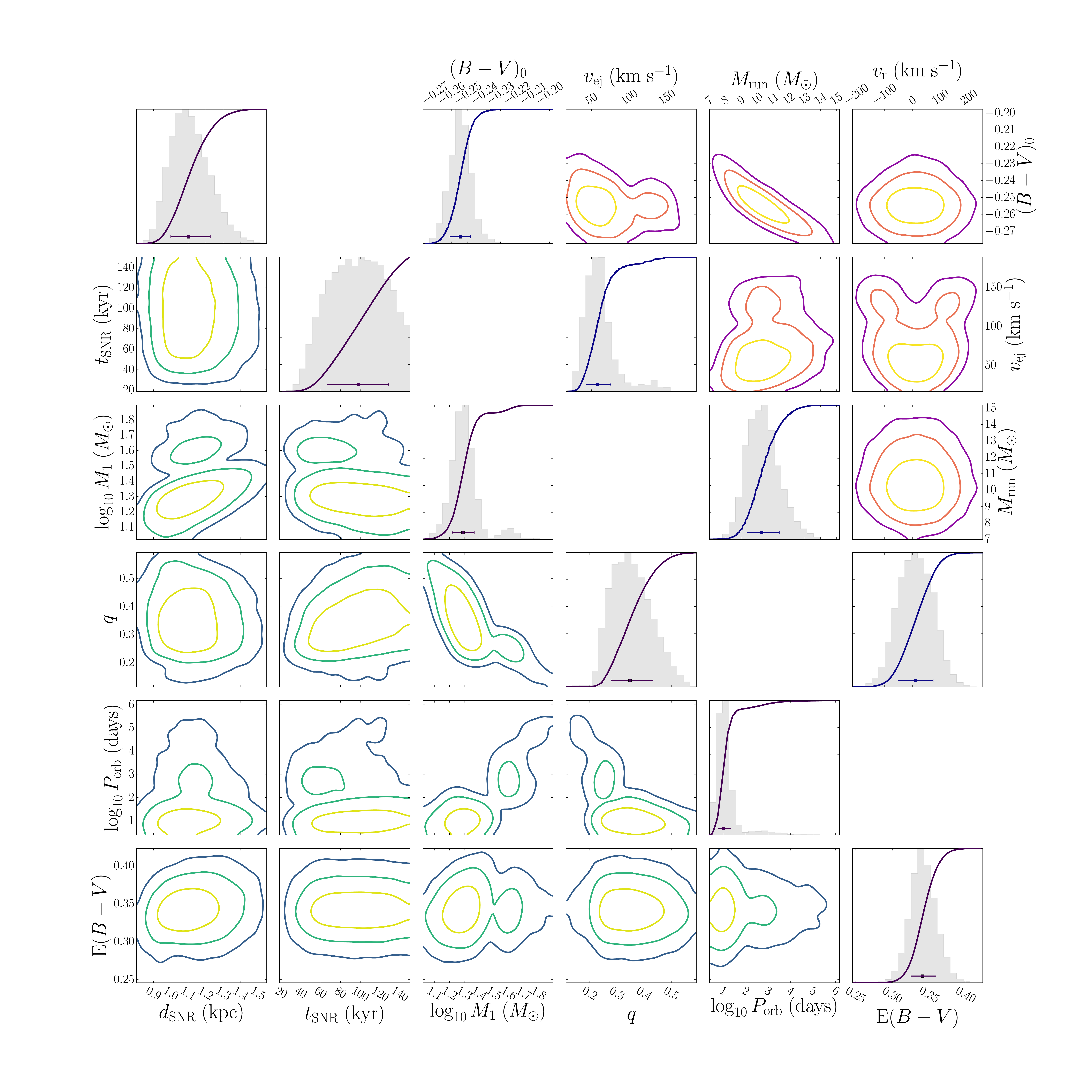}
	\caption{As Fig. \ref{fig:marginalG074.0} but for the
		candidate HD 37424 in G180.0$-01.7$.}
	\label{fig:marginalG180.0}
\end{figure*}

\begin{figure*}[h]
	\includegraphics[scale=0.28,trim = 55mm 45mm 40mm 40mm, clip]{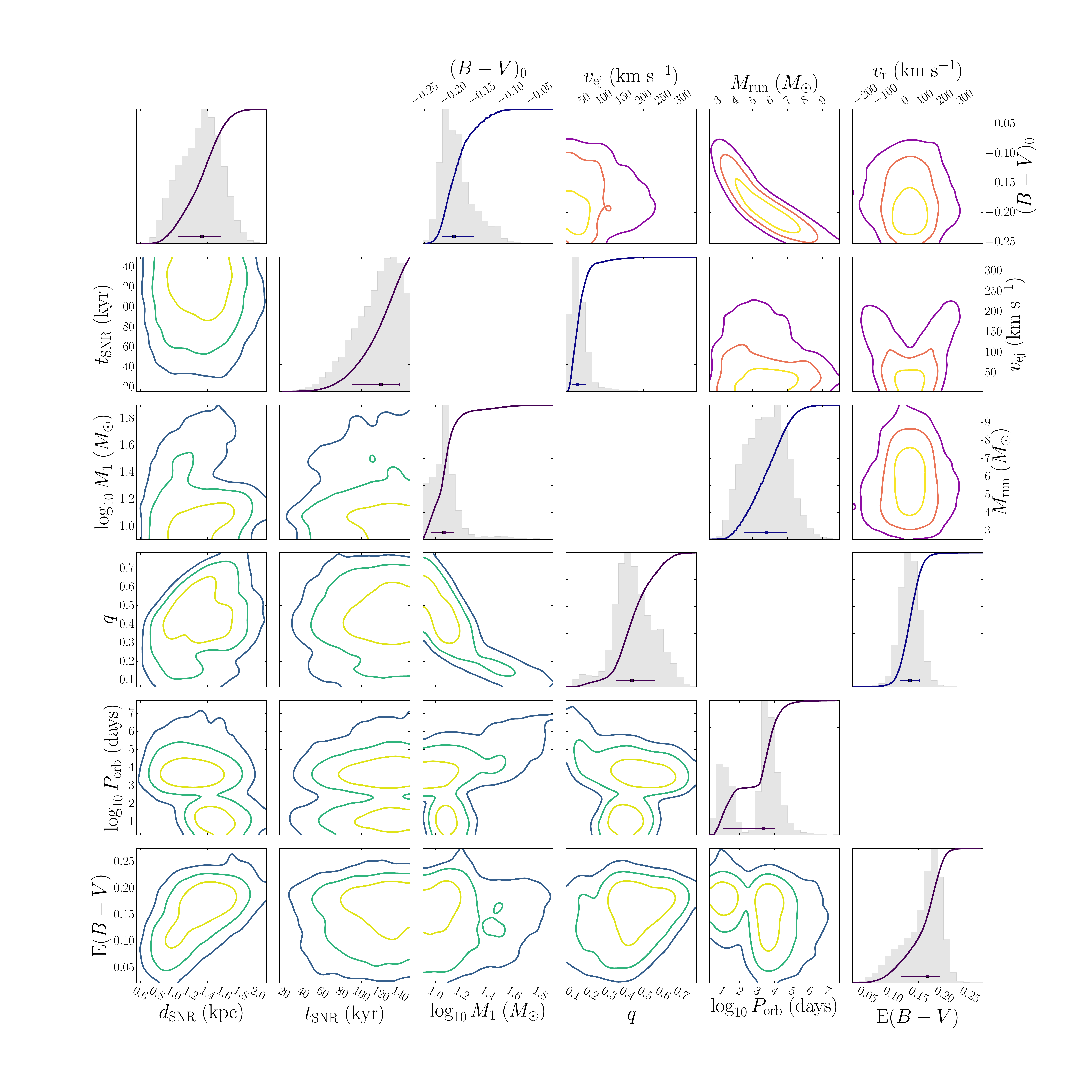}
	\caption{As Fig. \ref{fig:marginalG074.0} but for the
          candidate HD261393 in G205.5$+00.5$.}
	\label{fig:marginalG205.5}
\end{figure*}

\subsection{Individual candidates}
\label{sec:indivcand}

\textbf{TYC 2688-1556-1} The star TYC 2688-1556-1 in G074.0$-08.5$ (Cygnus Loop) has
no known references in the literature. It is a relatively high proper-motion
star with $(\mu_{\alpha
  *},\mu_{\delta})=(3.92\pm0.83,-21.03\pm1.25)\;\mathrm{mas}\;\mathrm{yr}^{-1}$
reported in TGAS. The colour and magnitude of this star in the
TGAS/APASS cross-match suggest this star is likely A type,
which agrees with the posterior for the current mass of the
runaway of $1.73\pm0.13\;M_{\odot}$. The posterior for the
ejection velocity includes a second mode which corresponds to the clump of stars at $v_{\mathrm{ej}}=700\;\mathrm{km}\;\mathrm{s}^{-1}$ in Figure \ref{fig:bvvej}. Runaways in this
region of $(B{-}V)_0\textbf{--}v_{\mathrm{ej}}$ space have undergone
significant mass exchange with the primary and will have had a common
envelope phase. This mass exchange shrinks the orbit of the binary which increases the orbital velocity and is the origin of the high velocity of these stars. If this mode is the true origin of TYC 2688-1556-1,
then the star is predicted to have lost several solar masses of
material, having started off at around $6\;M_{\odot}$ and ended with around $2\;M_{\odot}$. In
this case the star may be chemically peculiar. A more prominent
observable of this channel is that it would predict a heliocentric
radial velocity around $+ 600\;\mathrm{km}\;\mathrm{s}^{-1}$ or $- 600\;\mathrm{km}\;\mathrm{s}^{-1}$, where
the uncertainty is due to the degeneracy in whether the star is moving towards or
away from us. Looking at Figure \ref{fig:marginalG074.0}, this
degeneracy appears as a `v'-shaped contour in the
$v_{\mathrm{r}}\text{--}v_{\mathrm{ej}}$ plot. If the star is from
this mode, it is likely unbound from the Milky Way. The
covariance in the most probable mode between $M_1$ and $q$ is
simply the relationship $M_2=qM_1=\mathrm{const}$. This covariance is
interpreted as there being minimal mass transfer in the binary system
so that the mass of the runaway now is approximately the mass it was
born with. The secondary mode is clearly visible as lying off this
relationship.

\textbf{BD+50 3188} The B-type star BD+50 3188 in G089.0$+04.7$
exhibits emission lines in its spectra and so is classed as a Be star
\citep[most recently studied
  by][]{chojnowski_high-resolution_2015}. The emission lines in Be
stars are thought to originate from a low-latitude disk or ring-like
envelope \citep{kogure_astrophysics_2007}, which in the case of BD+50
3188 is measured to be rotating at $138\;\mathrm{km}\;\mathrm{s}^{-1}$
\citep{chojnowski_high-resolution_2015}. Be stars are also
characterised by rapid rotation, which can be close to their break-up
speed, and this is thought to be related to their formation mechanism
\citep{kogure_astrophysics_2007}. Be stars are observed in both single
and binary systems. There are plausible formation mechanisms in the
literature which can produce Be stars that are single or binary
\citep{kogure_astrophysics_2007}. A Be star can be formed if it is the
mass gaining component in a binary in the Roche-lobe Overflow (RLOF)
phase \citep[see][]{harmanec_emission-line_1987} where the emission-lines originate in the accretion disk formed of material lost by the Roche-lobe filling companion. \citet{pols_formation_1991} argues that the duration of the RLOF phase, and hence the lifetime of the accretion disk,
is not sufficiently long to explain the high fraction of B type stars
which are Be stars. \citet{pols_formation_1991} instead
propose the post-mass-transfer model where most Be stars are in systems after the end of RLOF. During RLOF the mass-gaining component is spun up by the angular momentum of the accreted mass. If the mass gainer is rotating at close to break-up by
the end of the RLOF phase we may see emission-lines from a decretion disk around the equator of the star. Shortly after the RLOF phase, the mass loser in such a system
detonates as a supernova which may unbind the system and produce
a runaway Be star
\citep{kogure_astrophysics_2007}. \citet{rinehart_single_2000} used
proper-motions from Hipparcos to compare the velocity distributions of
B and Be type stars and, finding that they were consistent to within
$1\sigma$, argued that they were not consistent with most Be type
stars being runaways. \citet{berger_search_2001} performed a similar
analysis but with the inclusion of radial velocities and found instead
that around $7\%$ of Be stars have large peculiar
velocities. \citet{berger_search_2001} points out that the runaway
fraction among all B stars is about $2\%$ and thus this result alone
supports a runaway origin for at least some Be
stars. \citet{berger_search_2001} goes on to argue that the fraction
of Be stars which have been spun up by binary interaction is
unknown, because some binaries will remain bound post-supernova (Be +
neutron star binary) and others will either be low-mass or experience
extreme mass-loss and bypass the supernova altogether (Be + Helium
star or Be + white dwarf binaries). In more recent work, \citet{de_mink_rotation_2013} modelled massive binary stars and found that it is possible for all early-type Be stars to originate in binaries through mass transfer and mergers. \citet{rivinius_classical_2013} review the origin and physics of Be stars and conclude that the emission-lines in a majority of the systems are due to a decretion disk around a rapidly rotating star, however they conclude that binarity is not a widespread mechanism because the statistical properties of B and Be binaries appear to be identical \citep{abt_binaries_1978,oudmaijer_binary_2010}. BD+50 3188 is the only Be star
within a $2\degr$ radius of G089.0$+04.7$ and is only
$2.4\;\mathrm{arcmin}$ from the centre. That it is a Be star with no
known binary companion which is spatially co-located with the SNR
lends circumstantial evidence to it being the runaway companion of
G089.0$+04.7$.

\textbf{HD 37424} This star is our most likely candidate with a Bayes factor $K$ of $2\ln K=17.72\pm0.13$. A connection between this star and the SNR G180.0$-01.7$ was previously drawn by \citet{dincel_discovery_2015} who used the kinematics of the star and the associated central compact object PSR J0538+2817 to show both were in the same location $30\pm4\;\mathrm{kyr}$ ago. \citet{dincel_discovery_2015} estimated that this star has spectral type $\mathrm{B}0.5\mathrm{V}\pm0.5$ and a mass around $13\;M_{\odot}$, while our method found $M_{\mathrm{run}}=10.38\pm1.04\;M_{\odot}$. \citet{dincel_discovery_2015} used this mass and the lack of nearby O-type stars to argue that the progenitor primary must have a mass that is at most $20\text{--}25\;M_{\odot}$, with the possibility that the system may have been a twin binary. The most likely mode in our posterior (Fig. \ref{fig:marginalG180.0}) corresponds to a scenario where the initial masses in the binary were $M_1=20\pm5\;M_{\odot}$ and $M_2=7\pm2\;M_{\odot}$. Our favoured initial primary mass is consistent with the lack of O-type stars while we find that the secondary has actually increased in mass because of mass transfer from the primary to the companion. The possibility of a twin progenitor binary is strongly excluded under our model.

Similarly to Section \ref{sec:simple} we took $B{-}V=0.073\pm0.025$ from \citet{dincel_discovery_2015} because HD 37424 is one of the five stars in G180.0$-01.7$ without APASS photometry. We were motivated to investigate this star despite it not having APASS photometry because it had been previously suggested to be the runaway companion.

\textbf{HD 261393} The star HD 261393 in G205.5$+00.5$ is given a
spectral type of B5V by \citet{voroshilov_catalog_1985} who also
assigned it membership of NGC 2244, an open cluster at the centre of
the Rosette Nebula. However, HD 261393 is $2\fdg5$ from the
centre of the Rosette Nebula (Fig. \ref{fig:rosette}), so it is
more likely to be a member of the adjoining Monoceros OB2 association
which extends to the east and north-east by several
degrees. \citet{odegard_decameter_1986} established that the Monoceros
Loop is within the Mon OB2 association and is interacting with, and
lies behind, the Rosette Nebula. This conclusion was supported by
later work (see \citealp{xiao_radio_2012} for a review).
\citet{martins_quantitative_2012} modelled the stellar properties of
ten O type stars in NGC 2244 and the surrounding Monoceros OB2
association and found that the age of the stars is in the range
$1\text{--}5\;\mathrm{Myr}$. In order for HD 261393 to be a runaway
with an age less than $5\;\mathrm{Myr}$, our model would require the
primary of the progenitor binary to be at least
$40\;M_{\odot}$. In the posterior shown in Figure
\ref{fig:marginalG205.5} a primary of this mass would lie between the
$2$ and $3\sigma$ contours. This extra constraint would decrease the
Bayesian evidence for a runaway origin and may be enough to result in
the background being more favourable. A similar line of reasoning for
the mass of the primary was put forward by \citet{gebel_nature_1972}
who argued that the minimum possible mass of the progenitor must exceed the $25\;M_{\odot}$ mass of the most massive O
star in the SNR. The models used by \citet{martins_quantitative_2012} to estimate the age of Mon OB2 did not include the possibility of rejuvenation by mass transfer or merger in binaries which can result in an underestimated age of OB associations (e.g. \citealp{schneider_ages_2014} used binary evolution simulations to predict that the $9\pm3$ and $8\pm3$ most massive stars in the Arches and Quintuplet star clusters respectively are likely merger products). Including the possibility of binary evolution would increase the estimated age of the Mon OB2 association and thus decrease the tension with our model. \citet{gebel_nature_1972} further speculated that the
B type star HD 258982 might be the associated runaway star because it
is the only B type star observed at that time in the SNR which
displays the \ion{CaK}{} absorption line at the
$16\;\mathrm{km}\;\mathrm{s}^{-1}$ of the expanding SNR shell. HD
258982 is around $1\fdg5$ away from the geometric centre of
the Monoceros Loop and the proper motion of this star had not been
measured at the time of \citet{gebel_nature_1972}. In TGAS, this star
has a measured proper motion of around
$3\;\mathrm{mas}\;\mathrm{yr}^{-1}$ meaning that the star can have
travelled at most $0\fdg1$ in the
$150\;\mathrm{kyr}$ age of the SNR and is effectively ruled
out as a possible candidate.

\section{Conclusions}
\label{sec:conclusions}

We have used two methods to search for and quantify the significance
of runaway former companions of the progenitors of nearby SNRs. The first method used
kinematics from the Tycho-{\it Gaia} astrometric solution (TGAS) to
find the star most likely to have been spatially coincident with the
SNR centre in the past $150\;\mathrm{kyr}$ and further filtered those
candidates based on their $B{-}V$ colour. This filtering was done to
select likely OB stars. The second method is more elaborate and was
designed to make full use of the available photometry, to incorporate
3D dustmaps, to be explicit about our expectation that most but not
all runaways are OB type, and to be statistically rigorous. This
Bayesian method has the advantage that it constrains the properties of
both the progenitor binary and the present day runaway.

Both methods returned four candidates and reassuringly three of those were in common. These are TYC 2688-1556-1 in G074.0$-08.5$, BD+50 3188
in G089.0$+04.7$ and HD 37424 in G180.0$-01.7$. The remaining candidate
from the kinematic method is TYC 4280-562-1 in G114.3$+00.3$ which has
$2\ln K=-4.69\pm0.14$ in the Bayesian method and thus is the seventh
most likely runaway in this SNR. The remaining candidate from the
Bayesian method is HD 261393 in G205.5$+00.5$, which was ranked fourth
in this SNR by the kinematic method.

Three of the candidates proposed by our Bayesian method are new, while
HD 37424 was previously suggested by \citet{dincel_discovery_2015}. It
is reassuring that this star was picked out by both methods and was
already in the literature. It has a Bayes factor $K$ of $2\ln
K=17.72\pm0.13$, which makes it a very strong candidate. The posterior
suggests that this star may have gained several solar masses from the
primary prior to the supernova. The best of our new candidates is
BD+50 3188. This is a Be star which can be explained by the star
being spun up by mass transfer from the primary prior to the
supernova. It is also the only Be star within several degrees of this
SNR and is only $2.4\;\mathrm{arcmin}$ from the geometric centre. If
TYC 2688-1556-1 is the runaway companion of G074.0$-08.5$ then it is
likely to be an A type. There is a second mode in the posterior for
TYC 2688-1556-1 which would correspond to this star having mass
transferred onto its primary. It predicts that this star may be
chemically peculiar and have a velocity greater than
$600\;\mathrm{km}\;\mathrm{s}^{-1}$, making it a hypervelocity
star. The final candidate from the Bayesian method is HD 261393. It is
possible that the progenitor of the Monoceros Loop is part of the
recent burst of star formation that has occurred in the Mon OB2
association over the last $1\text{--}5\;\mathrm{Myr}$. If this is
true, then this extra constraint may mean HD 261393 is more likely to
be a background star.

The method that \citet{dincel_discovery_2015} used to propose HD 37424
as a candidate was based on a coincident spatial location with the
pulsar in the past and thus is independent from our method which
relates the star to the properties of the SNR. One advantage of our
method is that it does not require there to be a known associated
pulsar. Our Bayesian method could be altered to include stellar radial
velocities and pulsar properties. The radial velocities would be an
additional constraint on the model, the pulsar parallax could provide
a more accurate distance to the SNR, and the pulsar proper motion
combined with a time since the SN would set the location of the
progenitor binary at the time of the SN. {\sc Gaia} is aiming to provide radial velocities for a bright subset of the main photometric and astrometric sample. It is estimated that for a B1V star with apparent magnitude $V=11.3$ the end-of-survey error on the radial velocity\footnote{\url{https://www.cosmos.esa.int/web/gaia/science-performance}} will be $15\;\mathrm{km}\;\mathrm{s}^{-1}$ which is sufficiently precise for tight constraints to be placed on runaway candidates.

A requirement of our Bayesian framework is the probability of a SNR to
have a runaway companion. Accounting for single stars, merging stars,
binaries that remain bound post-supernova and runaways that themselves
go supernova, we find that one third of core-collapse SNRs should have
a runaway companion. In agreement with this result, we find three
runaway candidates from the ten SNRs considered.

As mentioned previously, \citet{kochanek_cas_2017} ruled out runaway companions of the Crab, Cas A and SN 1987A SNRs with initial mass ratios $q\gtrsim0.1$. Including this null result for these three SNRs does not change our conclusion that the number of runaway candidates is consistent with the expected number of runaways, but if our two weaker candidates (TYC 2688-1556-1 and HD 261393) are subsequently ruled out a significant tension could arise. The SNRs considered by \citet{kochanek_cas_2017} are all younger ($t_{\mathrm{SNR}}<1\;\mathrm{kyr}$) and more distant ($d_{\mathrm{SNR}}>2\;\mathrm{kpc}$) than our SNR sample, making the two works complementary. The advantage of considering young SNRs is that a runaway companion is constrained to be much nearer to the centre of the SNR which limits the region that must be searched. The main disadvantage is the lack of parallaxes for distant stars which makes it harder to exclude candidates because of the degeneracy between distance, reddening and photometry. In terms of method \citet{kochanek_cas_2017} used PARSEC isochrones to carry out a pseudo-Bayesian fit to the photometry of each star while accounting for the distance and extinction to the SNR, which we would categorize as a middle ground between our simple and fully Bayesian approaches. \citet{kochanek_cas_2017} noted that a full simulation of binary evolution was beyond the scope of their work. It is the integration of binary evolution with a fully Bayesian method which is the main advance of this work. Future \emph{Gaia} data releases will allow our fully Bayesian method to be applied to both the Crab and Cas A SNRs.

{\it Gaia} Data Release 2 (DR2) will contain positions, parallaxes,
proper-motions and $G$, $G_{\mathrm{BP}}$ and $G_{\mathrm{RP}}$ for
over a billion stars. This dataset is the reason for constructing our
Bayesian framework. The millimagnitude precision of the photometry
will remove poorly measured stars as contaminants, while the
milliarcsecond precision of the parallaxes will remove high
proper-motion foreground stars. The final {\it Gaia} data release aims
to be complete down to $G\approx20.5$ and at that completeness we will
be able to test the existence of a runaway companion for all the nearby SNRs.

Future spectroscopic observations of BD+50 3188, TYC 2688-1556-1 and
HD 261393 will test whether they truly are SN companions, allowing
them to be used to test binary star evolution. With {\it Gaia} DR2 in
early 2018, our Bayesian framework provides a sharp set of tools that
will allow us to find any runaways there are to find.

\begin{acknowledgements}
We thank Sergey Koposov, Vasily Belokurov, Jason Sanders and other members of the
Streams group at the Institute of Astronomy in Cambridge for comments
while this work was in progress. We also thank Gerry Gilmore for early discussions on this work. We are especially appreciative of Mathieu Renzo, Simon
Stevenson, Manos Zapartas and the many other authors cited above who
have contributed to the development of {\sc binary\_c}. DPB is grateful to the Science and
Technology Facilities Council (STFC) for providing PhD funding. MF is supported by a Royal Society - Science Foundation Ireland University Research Fellowship. This work was partly supported by the European Union FP7 programme through ERC grant number 320360. RGI thanks the STFC for funding his Rutherford fellowship under grant ST/L003910/1 and Churchill College, Cambridge for his fellowship. This
work has made use of data from the European Space Agency (ESA) mission
{\it Gaia} (\url{https://www.cosmos.esa.int/gaia}), processed by the
{\it Gaia} Data Processing and Analysis Consortium (DPAC,
\url{https://www.cosmos.esa.int/web/gaia/dpac/consortium}). Funding
for the DPAC has been provided by national institutions, in particular
the institutions participating in the {\it Gaia} Multilateral
Agreement. This research has made use of the APASS database, located
at the AAVSO web site. Funding for APASS has been provided by the
Robert Martin Ayers Sciences Fund. Figures \ref{fig:bvvej} and \ref{fig:ebv} made use of the {\sc cubehelix} colour scheme \citep{green_colour_2011}.
\end{acknowledgements}

\bibliographystyle{aa} 
\bibliography{references} 

\begin{appendix} 

\section{The Local Supernova Remnants (SNRs)}
\label{sec:boutique}

Our strategy to obtain distances is simple. We begin with the list of
SNRs with known distances in the literature which
\cite{pavlovic_updated_2014} used to calibrate their
$\mathit{\Sigma}\text{--}D$ relation. We check the source cited by
\cite{pavlovic_updated_2014} for each measurement. We then conduct our
own literature search to see if there are more recent distances
available, starting with the SNR catalogues of
\citet{green_catalogue_2014} and \citet{ferrand_census_2012}. Some
distances for SNRs in the \citet{green_catalogue_2014} catalogue are
given in the more detailed online version\footnote{Green D. A., 2014,
  `A Catalogue of Galactic Supernova Remnants (2014 May version)',
  Cavendish Laboratory, Cambridge, United Kingdom (available at
  \url{http://www.mrao.cam.ac.uk/surveys/snrs/}).}. Below we discuss
our arguments for the chosen distance and age used for each SNR.
However, we emphasise that the distance is not a major factor in our
method since we constrain the runaways to lie on the main sequence and
so constrain the distance to a narrow range. For the SNRs where
there is no error attached to the best distance estimate, we assume a
nominal 50\% error.

\begin{enumerate}
  
\item{G065.3$+$5.7} : \citet{boumis_kinematics_2004} combined an expansion
  velocity measurement of $155\;\mathrm{km}\;\mathrm{s}^{-1}$ with a
  proper motion in the optical of $2.1\pm0.4\;\mathrm{arcsec}$ in 48
  years to derive a distance $0.77\pm0.2\;\mathrm{kpc}$.

 \medskip

\item {G069.0$+$2.7 (CTB 80)} : A commonly cited
  distance estimate for G069.0$+02.7$ is $2\;\mathrm{kpc}$ from
  \citet{koo_detection_1990}, however in the original paper the
  estimate is given in the form $2d_2\;\mathrm{kpc}$ where $d_2$ is a
  scaling factor of order unity. \citet{koo_interaction_1993}
  constrained this parameter to $1.0\pm0.3$. \citet{leahy_radio_2012} bound the distance to the range
  $1.1\text{--}2.1\;\mathrm{kpc}$ and pick a nominal distance of
  $1.5\;\mathrm{kpc}$. We assume a distance $1.5\pm0.5\;\mathrm{kpc}$, where we
  have added a nominal $0.5\;\mathrm{kpc}$ error.

  \medskip
  
\item{G074.0$-$8.5 (Cygnus Loop)} : \citet{blair_hubble_2005} used a
  measured shock velocity of $155\;\mathrm{km}\;\mathrm{s}^{-1}$ with
  HST proper motion of
  $0.070\pm0.008\;\mathrm{arcsec}\;\mathrm{yr}^{-1}$ to derive a
  distance $0.54_{-0.08}^{+0.10}\;\mathrm{kpc}$.

  \medskip

\item{G089.0$+$4.7 (HB 21)} : There are two competing distance estimates in the
  literature. \citet{tatematsu_interaction_1990} arrived at a distance
  estimate by establishing an interaction with molecular clouds in the
  Cyg OB7 association and then taking the distance of that association
  $0.80\pm0.07\;\mathrm{kpc}$ \citep{humphreys_studies_1978} to be the
  distance of the SNR. Note, however, that a more recent distance
  estimate of Cyg OB7 using Hipparcos parallaxes
  \citep{esa_hipparcos_1997} gives the distance $0.6\;\mathrm{kpc}$
  \citep{melnik_kinematics_2009}. \citet{byun_interaction_2006}
  discussed the link between HB 21 and molecular clouds in Cyg OB7 and
  argued that, while there were morphological similarities, there was
  no direct evidence for the
  association. \citet{byun_interaction_2006} discusses other distance
  estimates in the literature and arrives at a distance estimate of
  $1.7\pm0.5\;\mathrm{kpc}$. One key argument used by
  \citet{byun_interaction_2006} is that the X-ray surface brightness
  of HB 21 is too faint for $0.8\;\mathrm{kpc}$ and that it must be
  beyond $1.6\;\mathrm{kpc}$ \citep{yoshita_ph.d._2001}. Updating the
  distance of Cyg OB7 using the Hipparcos parallaxes increases this
  tension and favours the distance $1.7\pm0.5\;\mathrm{kpc}$. We assume that HB 21 lies at
  $1.7\pm0.5\;\mathrm{kpc}$.

  \medskip
  
\item{G093.7$-$0.2 (CTB 104A, DA 551)} :
  \citet{uyaniker_supernova_2002} calculated a distance of
  $1.5\pm0.2\;\mathrm{kpc}$ based on the kinematics of \ion{H}{I}
  features associated with the remnant. We assume this distance
  estimate.

%

\medskip

\item{G114.3$+$0.3} : The most recent estimate of $0.7\;\mathrm{kpc}$ is based on
  association with \ion{H}{I} emission features
  \citep{yar-uyaniker_distance_2004}. We assume a distance estimate of
  $0.70\pm0.35\;\mathrm{kpc}$ with a nominal 50\% error.

  \medskip
  
\item{G119.5$+$10.2 (CTA 1)} : \citet{pineault_supernova_1993} calculated a distance of $1.4\pm0.3\;\mathrm{kpc}$ based
  on an association with an \ion{H}{I} shell.

  \medskip
  
\item{G160.9$+$2.6 (HB 9)} : \citet{leahy_radio_2007}
  estimated the distance at $0.8\pm0.4\;\mathrm{kpc}$ using \ion{H}{I}
  absorption.

  \medskip
  
\item {G180.0$-$1.7 (S147)} : G180.0$-01.7$ is notable for being
  a nearby SNR with a convincing runaway candidate
  \citep{dincel_discovery_2015}.   \citet{sallmen_intermediate-velocity_2004} noted that
  HD 36665 at $880\;\mathrm{pc}$ \citep[photometric distance
    from][]{phillips_high-velocity_1981} and HD 37318 at
  $1380\;\mathrm{pc}$ had absorption lines at a similar velocity to
  the expansion of the SNR shell, while HD 37367 at
  $361_{-85}^{150}\;\mathrm{pc}$ \citep[parallax
    from][]{esa_hipparcos_1997} did not have such lines. \citet{sallmen_intermediate-velocity_2004} estimated a distance of $0.62\;\mathrm{kpc}$ based on the SNR lying in between HD 37367 and HD 36665. We were
  unable to locate the original source for the distance estimate of HD
  37318. \citet{dincel_discovery_2015} argued that the likely
  association of G180.0$-01.7$ with the pulsar PSR J0538+2817 makes the
  most accurate distance $1.30_{-0.16}^{+0.22}\;\mathrm{kpc}$
  \citep[parallax measurement by][]{chatterjee_precision_2009}. The
  tension between the distance derived by looking for stars in front
  and behind the supernova shell and the distance obtained from the
  parallax of the associated pulsar may be relieved by more accurate
  distance measurements from the second {\it Gaia} data release. We
  assume the distance estimate $1.30_{-0.16}^{+0.22}\;\mathrm{kpc}$.

%

    \medskip
    
\item{G205.5$+$0.5 (Monoceros Loop)} :
  There are two distances in the literature. A distance of $0.6\;\mathrm{kpc}$ based on the mean optical velocity and a
  distance of $1.6\;\mathrm{kpc}$ from \citet{xiao_radio_2012}. We
  assume a distance of $1.2\pm0.6\;\mathrm{kpc}$, where we take a
  nominal 50\% error.

  \end{enumerate}
  
\section{Implementation of {\sc MultiNest}}
\label{sec:multinest}

{\sc MultiNest} explores the parameter space of a model by choosing
new samples from within an ellipse containing the current
samples. This sampling requires that the prior be expressible as a
uniform distribution on the unit hypercube. To encode non-trivial
distributions, these $N$ random variables distributed as $U(0,1)$ must
be transformed into the parameter space. Through this procedure, the
prior is automatically normalised. For independent random variables,
this is a simple application of inverse transform sampling. However,
for dependent variables $\boldsymbol{x}$, the simplest course is usually
to move the prior $f(\boldsymbol{x})$ to the likelihood function and use a
uniform prior over the entire permitted parameter space for each
variable when doing the transform. A correction must then be applied
to remove the normalisation that {\sc MultiNest} has applied in the
prior. This becomes non-trivial if there is any area of the parameter
space where the likelihood function returns a negligibly small number,
since {\sc MultiNest} treats that area of the parameter space as
invalid and so renormalises the parameter space to exclude it. This
behaviour is problematic if the likelihood is zero over large parts of
the parameter space because it means that the calculated evidence is
wrong. The reason for this behaviour is that it allows the user to
encode constraints between variables in the prior. One physical
example is if the radii $R_1$ and $R_2$ of two stars in a binary
system are not constrained, but the separation $a$ is known. This can
be encoded as $R_1+R_2<a$ and implemented by returning zero in the
likelihood if the constraint is broken. {\sc MultiNest} then
renormalises the prior to exclude those regions that break the
constraint. This behaviour is one way to have a uniform distribution
over an arbitrarily complicated support. To sidestep this behaviour we
instead return $ue^{-10^{20}}$ where $u\sim U(1,1.0001)$, since this
is both a negligibly small number and above the default threshold for
{\sc MultiNest} to ignore. The reason for including the random
variation is that in Nested Sampling the likelihood of the points is
sorted as part of the algorithm and the case where two points have the
same likelihood is important. \citet{skilling_nested_2006} mentions
that it is necessary to resolve ties between points of equal
likelihood and that this can be achieved by adding a small random
number of sufficient precision that repeats are unlikely.

\end{appendix}
\end{document}